\documentclass[sigconf, nonacm]{acmart}
\usepackage{algorithm}
\usepackage{algorithmicx}
\usepackage{algpseudocode}
\usepackage{amsmath} 
\usepackage{mdframed}% For mathematical symbols
\usepackage{booktabs}
\usepackage{multirow}
\usepackage{graphicx}
\usepackage{xcolor}
\usepackage{graphicx}

\usepackage{listings}
\usepackage{cleveref}
\usepackage{adjustbox}

\AtBeginDocument{%
  }

\usepackage{enumitem}

\newcommand{\ourmodel}{{\textsc{FinSage}}}

\begin{document}


\title{\includegraphics[height=25pt]{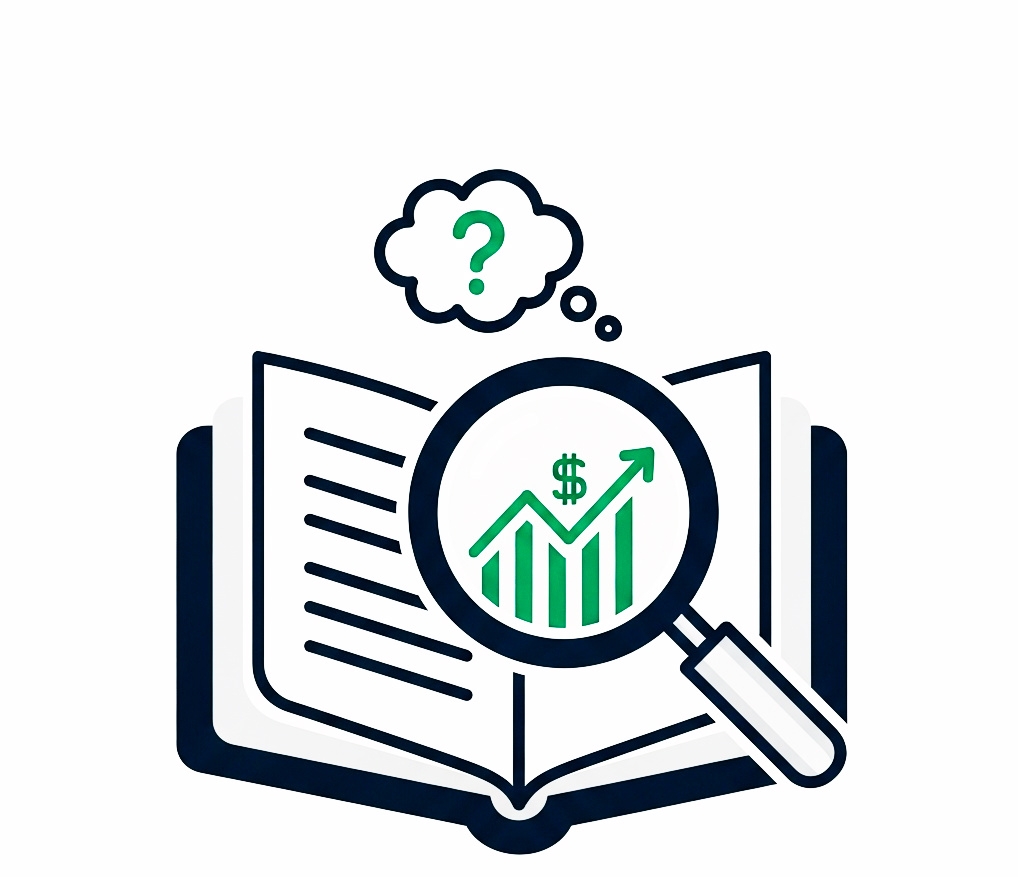}\ourmodel{}: A Multi-aspect RAG System for Financial Filings Question Answering}

\author{
    Xinyu Wang\textsuperscript{1,2 *},
    Jijun Chi\textsuperscript{1,3 *},
    Zhenghan Tai\textsuperscript{1,3 *},
    Tung Sum Thomas Kwok\textsuperscript{1,4},
    Muzhi Li\textsuperscript{5}, \\
    Zhuhong Li\textsuperscript{1,6},
    Hailin He\textsuperscript{1},
    Yuchen Hua\textsuperscript{1,2},
    Peng Lu\textsuperscript{7},
    Suyuchen Wang\textsuperscript{7,8},
    Yihong Wu\textsuperscript{7}, \\
    Jerry Huang\textsuperscript{7,8},
    Jingrui Tian\textsuperscript{4},
    Fengran Mo\textsuperscript{7},
    Yufei Cui\textsuperscript{2,9 \textdagger},
    Ling Zhou\textsuperscript{10 \S}
}

\affiliation{
    \institution{\textsuperscript{1}SimpleWay.AI \quad
    \textsuperscript{2}McGill University \quad
    \textsuperscript{3}University of Toronto \quad
    \textsuperscript{4}University of California, Los Angeles}
    \institution{\textsuperscript{5}The Chinese University of Hong Kong \quad
    \textsuperscript{6}Duke University \quad
    \textsuperscript{7}Universit\'e de Montr\'eal \quad
    \textsuperscript{8}Mila - Quebec AI Institute}
    \institution{\textsuperscript{9}Noah's Ark Lab \quad
    \textsuperscript{10}CG Matrix Technology Limited}
    \country{}
}

\thanks{\textsuperscript{*}Core Contribution.
        \par\textsuperscript{\textdagger}Corresponding Author - yufei.cui@mail.mcgill.ca
        \par\textsuperscript{\S}Senior Author - ling.zhou@cgmatrix.com
        }

% \authornote{Core Contribution, \dagger \textnormal{ Corresponding Author}, \S \textnormal{ Senior Author}}

% \authornote{\textnormal{Core Contribution}, \dagger \textnormal{ Corresponding Author}, \S \textnormal{ Senior Author}}
% \authornote{\textnormal{Core Contribution}}
% \authornote{\textnormal{Corresponding Author}}
% \authornote{\textnormal{Senior Author}}
% \authornote{\texttt{\{xinyu.wang5, yufei.cui\}@mail.mcgill.ca}}
% \authornote{\texttt{\{ferris.chi, winfred.tai\}@mail.utoronto.ca}}
% \authornote{\texttt{ling.zhou@cgmatrix.com}}

% \email{xinyu.wang5@mail.mcgill.ca, ferris.chi@mail.utoronto.ca, \\ winfred.tai@mail.utoronto.ca, yufei.cui@mail.mcgill.ca, ling.zhou@cgmatrix.com}

\email{xinyu.wang5@mail.mcgill.ca}
\email{{ferris.chi, winfred.tai}@mail.utoronto.ca}
% \email{ling.zhou@cgmatrix.com}
% 

% \affiliation{%
%   \city{Toronto}
%   \state{ON}
%   \country{Canada}
% }

% \affiliation{%

% \affiliation{%

% \affiliation{%

% \affiliation{%

% \affiliation{%

% \affiliation{%

% \affiliation{%

\renewcommand{\shortauthors}{Xinyu et al.}

\begin{abstract}

Leveraging large language models in real-world settings often entails a need to utilize domain-specific data and tools in order to follow the complex regulations that need to be followed for acceptable use. Within financial sectors, modern enterprises increasingly rely on Retrieval-Augmented Generation (RAG) systems to address complex compliance requirements in financial document workflows.
However, existing solutions struggle to account for the inherent heterogeneity of data (e.g., text, tables, diagrams) and evolving nature of regulatory standards used in financial filings, leading to compromised accuracy in critical information extraction.
We propose the \ourmodel{} framework as a solution, utilizing a multi-aspect RAG framework tailored for regulatory compliance analysis in multi-modal financial documents.
\ourmodel{} introduces three innovative components: (1) a multi-modal pre-processing pipeline that unifies diverse data formats and generates chunk-level metadata summaries, (2) a multi-path sparse-dense retrieval system augmented with query expansion (HyDE) and metadata-aware semantic search, and (3) a domain-specialized re-ranking module fine-tuned via Direct Preference Optimization (DPO) to prioritize compliance-critical content. Extensive experiments demonstrate that \ourmodel{} achieves an impressive recall of 92.51\% on 75 expert-curated questions derived from surpasses the best baseline method on the FinanceBench question answering datasets by 24.06\% in accuracy. Moreover, \ourmodel{} has been successfully deployed as financial question-answering agent in online meetings, where it has already served more than 1,200 people.

%Large language models, commonly used in commercial domains, fall short in processing complex multi-modal data 

%Large Language Models~(LLMs) and Retrieval-Augmented Generation~(RAG) frameworks have demonstrated remarkable capabilities in natural language processing, including applications in the financial domain. However, financial documents pose unique challenges due to their complexity, domain-specific terminology, and the critical need for factual accuracy. Existing models often struggle with hallucination, retrieval inefficiencies, and inadequate contextual understanding, limiting their reliability in high-stakes financial applications.

\end{abstract}






\maketitle
% \footnotetext[1]{Core Contribution.}
% \footnotetext[2]{Corresponding Author.}
% \footnotetext[4]{Senior Author.}

\vspace{0.4cm}
\begin{figure}[th]
  \centering
  \includegraphics[width=\linewidth]{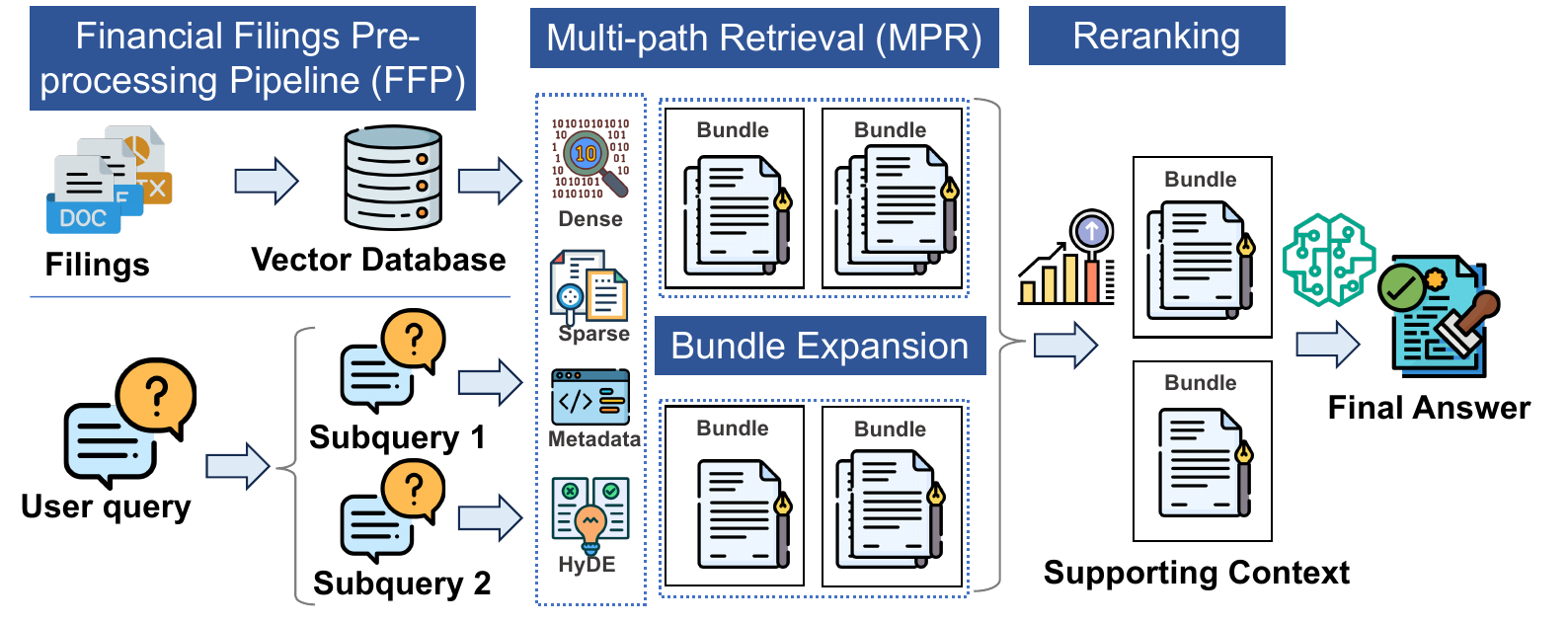}
  \caption{An overview of our proposed~\ourmodel{}.}
  \label{fig:overview-simp}
  \vspace{-0.4cm}
\end{figure}

\section{Introduction}
% \authorthomas

In the real world, various regulatory constraints define the manners in which tools and systems can be permissibly integrated into modern workflows, complicating the manners in which technologies such as large language models (LLMs) can be directly play a role in fulfilling specific tasks. One such domain is the financial sector, where adherence to regulations is essential to mitigate legal risk and maintain stakeholder confidence. However, financial regulations can evolve with time, presenting complex and significant challenges even for well-structured institutions with expert teams dedicated for this specific purpose~\citep{gs-sec, jpm-sec}.
% Financial filings are often lengthy, multi-modal, and nuanced~\citep{financial-filing, financial-filing-lengthy, cao2024characterizingmultimodallongformsummarization, xie2024openfinllmsopenmultimodallarge}, requiring meticulous analysis to extract key regulatory insights.
More practically, these challenges have led to a strong demand for a Question Answering (QA) system capable of efficiently retrieving and analyzing compliance-related information from financial filings~\citep{shah2024multidocumentfinancialquestionanswering}, with the hope of reducing manual labor and minimizing human error. Such systems often require the integration of retrieval-augmented generation (RAG), which enables machine learning-powered systems to directly search through databases to extract relevant information that can empower greater decision making.

However, while RAG systems have proven helpful for general LLM reliability and reasoning, integration into financial compliance tasks can be difficult due to inherent domain-related limitations. First, financial filings combine unstructured text (narrative disclosures), semi-structured data (tables, graphs), and contextual metadata - a multi-modal composition that conventional text-centric retrieval architectures fail to process cohesively, leading to fragmented or incomplete context representation~\citep{xie2024openfinllmsopenmultimodallarge, riedler2024textoptimizingragmultimodal, zhai2024selfadaptivemultimodalretrievalaugmentedgeneration, xia2024mmedragversatilemultimodalrag}. Next, while existing RAG systems predominantly employ dense retrieval~\citep{karpukhin2020dense, xiong2020approximate} or sparse lexical matching~\citep{robertson2009probabilistic}, these can fail due to the need for domain-specific fine-tuning or an inability to capture implicit regulatory relationships~\citep{JimenoYepes2024FinancialRC, yuan2024hybridragcomprehensiveenhancement, sultania2024domainspecific}.
%these approaches exhibit critical shortcomings in compliance-driven scenarios. Dense retrievers require domain-specific fine-tuning hampered by scarce labeled financial data and computational costs, whereas sparse methods lack the semantic awareness to capture implicit regulatory relationships (e.g., causal links between risk disclosures and mitigation strategies)~\citep{yepes2024financialreportchunkingeffective, yuan2024hybridragcomprehensiveenhancement, sultania2024domainspecificquestionansweringhybrid}. 
% Finally, even when retrieval succeeds, ranking ambiguity arises when multiple candidate chunks exhibit partial relevance—a frequent occurrence in lengthy filings. Current reranking strategies often prioritize general semantic similarity over compliance-critical criteria, propagating errors into downstream generation and increasing hallucination risks
Finally, existing retrieval methods often rely on ranking relevant passages based on semantic similarity rather than domain-specific criteria, leading to the critical risk of faulty or incomplete reasoning~\citep{barnett2024sevenfailurepointsengineering, besta2024multiheadragsolvingmultiaspect, agrawal2024mindfulragstudypointsfailure}. Such limitations collectively undermine the precision and accountability of automated compliance workflows, exposing enterprises to operational and regulatory hazards. 

As a solution to these challenges, we introduce \ourmodel{}, an end-to-end financial question-answering framework optimized for regulatory compliance that leverages the heterogeneous nature of financial data for better reasoning. As shown in \autoref{fig:overview-simp}, \ourmodel{} consists of three core components: (1) Financial Filings Pre-processing (FFP) pipeline, (2) Multi-path Retrieval (MPR) pipeline, and (3) Domain-Specialized Document Re-ranker, which collectively overcome the aforementioned challenges of multi-modal documents domain-agnostic retrieval, and ranking ambiguity.
First, the \textit{FFP module} tackles the multi-modal heterogeneity in financial documents by unifying text, tables, and graphs into a structured vector database through modality-specific encoders.
%To mitigate information redundancy, it employs coreference resolution and deduplication filters, while generating chunk-level metadata summaries that preserve regulatory context.
Next, the \textit{MPR module} augments traditional dense/sparse retrieval with two tailored domain-aware strategies, ensuring robust coverage of both explicit and implicit domain-specific relationships.
% Second, the MPR module augments traditional dense/sparse retrieval with two domain-aware strategies: (1) metadata-aware retrieval that leverages chunk summaries to answer high-level semantic queries (e.g., \textit{Identify Q3 sales trends discussed in Management Discussion and Analysis}), and (2) hypothesis-based retrieval (HyDE~\citep{gao2022precise, zhang2024exploring}), which expands queries using synthetic regulatory compliance hypotheses (e.g., inferring unstated causal relationships between risk factors). This hybrid approach ensures robust coverage of both explicit and implicit domain-specific relationships. 
Finally, the \textit{Document Re-ranker} addresses precision limitations through direct preference optimization (DPO)~\citep{zhang2024instructiontuninglargelanguage}, prioritizing chunks containing legally significant phrases (e.g., ``material weakness'' or ``non-compliance'') and suppressing irrelevant semantic matches. By jointly optimizing retrieval diversity and precision, \ourmodel{} establishes a new paradigm for compliance-critical financial QA systems. The experimental results demonstrate that \ourmodel{} significantly outperforms traditional RAG approaches in financial regulatory compliance tasks, offering a practical and scalable solution for financial institutions. In summary, our contributions are as follows:
\begin{itemize}[leftmargin=*]
    \item We present an end-to-end financial QA system that integrates pre-processing, retrieval, and re-ranking to provide a robust solution for regulatory compliance.
    \item We address key challenges in multi-modal financial document processing and domain-specific question answering, bridging the gap between traditional RAG models and regulatory compliance applications.
    \item We conduct extensive experiments and ablation studies to evaluate the effectiveness of our retrieval, re-ranking, and response generation mechanisms, analyzing the impact of different retrieval paths and re-ranking strategies.
\end{itemize}

\begin{figure*}[h!]
  \centering
  \includegraphics[width=\linewidth]{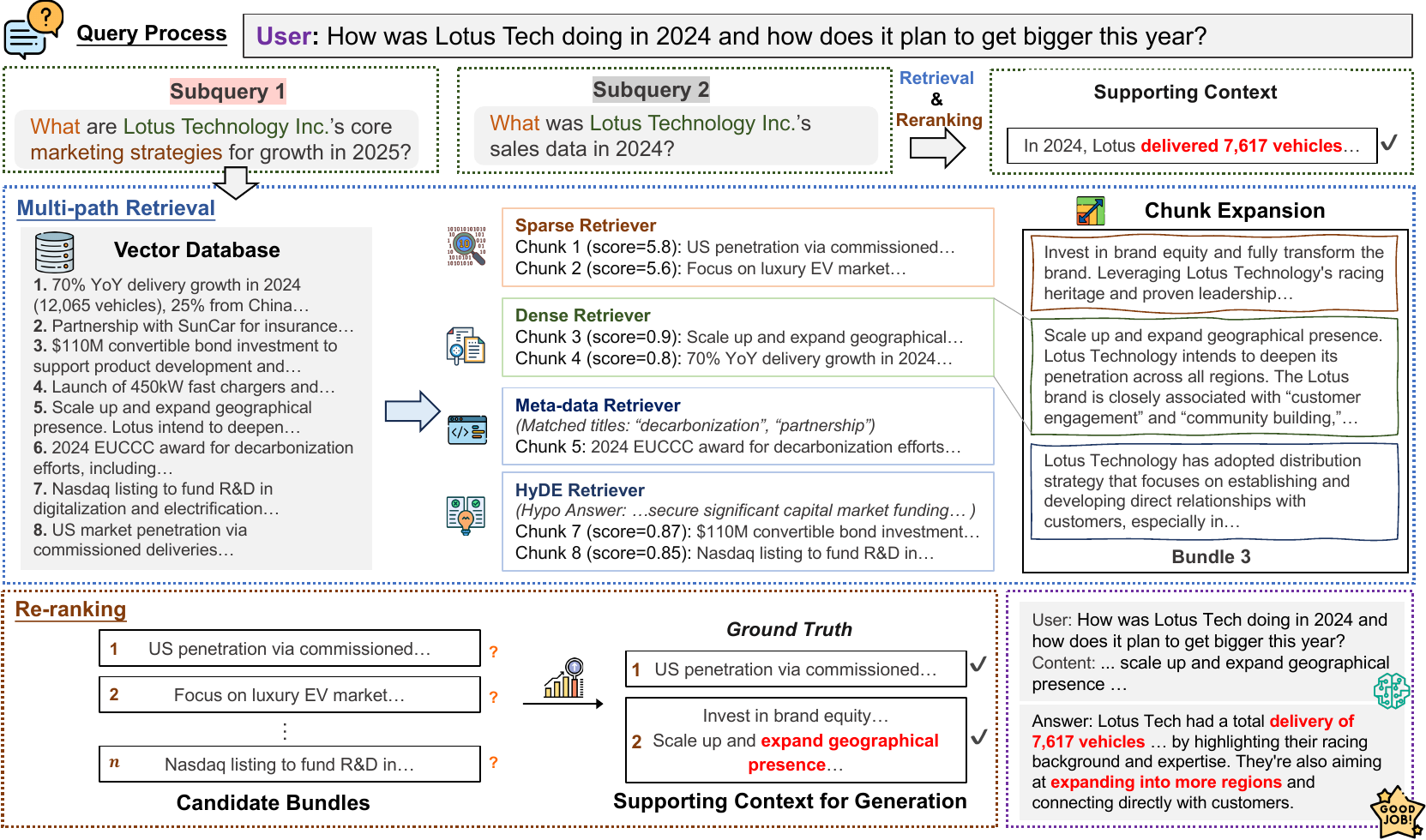}
  \caption{The complete \ourmodel{} RAG pipeline. The system processes the user's query and splits it into subqueries. For each subquery, multiple chunks are retrieved with multi-path retrieval, and these chunks are then expanded into candidate bundles by concatenating neighboring chunks. After re-ranking, the most relevant chunks and their bundles are included for answer generation.}
  \label{fig:overview}
\end{figure*}

\section{Related Work}

\subsection{Retrieval Augmented Generation}

Despite their strong empirical performance, large language models (LLMs) are inherently bottlenecked by their fixed parametric size, limiting their ability to store factual knowledge, which can affect their performance on specific tasks. Retrieval-Augmented Generation (RAG)~\citep{RAG2,ma2025thinkongraph} emerged as a solution by integrating external knowledge that the LLM could access through retrieval within an external database. A retrieval module first selects relevant documents from sources like news, academic papers, or social media, which are then combined with the input query and fed into a language model to generate a response. This approach leverages both the model's internal memory and the retrieved corpus~\citep{zhao2024longragdualperspectiveretrievalaugmentedgeneration}, ensuring more accurate and contextually grounded outputs.

Recent studies have explored various improvements, such as feedback tokens for relevance~\citep{asai2024selfrag}, self-referential knowledge acquisition~\citep{wang2023selfknowledge}, iterative self-feedback~\citep{liu2024raisf}, and domain-specific hierarchies~\citep{wang2024biorag}. In cases where initial queries lack sufficient detail, generative retrieval enhances information retrieval by generating queries or synthesizing passages beyond traditional keyword-based search~\citep{rajput2023recommendersystemsgenerativeretrieval, li2024matchinggenerationsurveygenerative, anonymous2024tabmeta}. Some studies further work to use multi-head RAG to extract information from diverse sources~\citep{besta2024multiheadragsolvingmultiaspect, xia2024mmedragversatilemultimodalrag}, aiming to maximize answer accuracy through quantity-based retrieval. Others focus on enhancing answer precision via quality retrieval, leveraging re-rankers~\citep{jacob2024drowningdocumentsconsequencesscaling, moreira2024enhancingqatextretrieval,SSET} to preserve high answer relevancy.

\subsection{Multi-Modal Retrieval}

Information is not limited to text, meaning that knowledge from various modalities can be used for improved reasoning. Some approaches, such as MuRAG~\citep{chen-etal-2022-murag} and generative retrieval frameworks~\citep{10.1609/aaai.v38i17.29837}, enhance retrieval by leveraging multi-modal memory, generative retrieval, and knowledge-guided decoding, improving performance in multi-modal QA tasks. Others, including VisRAG~\citep{yu2025visrag} and multi-modal RAG for industrial applications~\citep{riedler2024textoptimizingragmultimodal}, explore vision-language models to retain document structure and improve retrieval accuracy, particularly in industrial and document-heavy settings.

\subsection{Retrieval for Financial Data}

Despite recent advances, RAG-LLM systems remain underdeveloped for financial filings. \citet{islam2023financebench} introduced \textsc{FinanceBench}, a dataset specifically designed for financial document understanding. Building on this, \textsc{Unstructured}~\citep{JimenoYepes2024FinancialRC} explores different chunking strategies to enhance financial document processing. \citet{setty2024improvingretrievalragbased} proposes a RAG pipeline to facilitate retrieval in financial reports. However, existing approaches still face significant challenges: they struggle to handle complex multi-modal information—such as figures and tables—and their simple cross-encoder-based re-rankers often lack precision, negatively impacting retrieval performance.

\section{Methodology}

To reliably use RAG for financial document processing, the system must address three fundamental challenges: (1) handling multi-modal data such as tables, figures, and structured text; (2) extracting a comprehensive scope of relevant information from multiple sources; and (3) generating accurate and precise responses by effectively summarizing and reasoning over retrieved content. To tackle these challenges, we propose \ourmodel{}, an end-to-end financial QA system tailored for regulatory compliance.

This section introduces the three key components of~\ourmodel{}, each designed to address specific challenges in financial RAG (Figure \ref{fig:overview}). Section~\ref{sub: ffp} presents the \textbf{Financial Filings Pre-processing Pipeline (FFP)}, which transforms financial documents into a structured machine-readable format while enriching their semantic content. Section~\ref{sub:multi-path-retrieval-pipeline} details the \textbf{Multi-path Retrieval Pipeline (MPR)}, which enhances retrieval accuracy and contextual relevance by integrating multiple retrieval strategies. Section~\ref{sub: reranker} discusses the \textbf{Document Re-ranker}, which improves retrieval precision by refining and prioritizing the most relevant document chunks. 

% \textbf{Problem Definition.} 
% [Placeholder: The problem modeling and formalization will be introduced here.]

% Each of these components plays a crucial role in ensuring that \ourmodel{} effectively retrieves and processes financial data, providing reliable and contextually accurate responses. The subsequent sections provide a detailed discussion of their design and implementation.

\begin{figure*}[h] 
    \centering 
    \includegraphics[width=\linewidth]{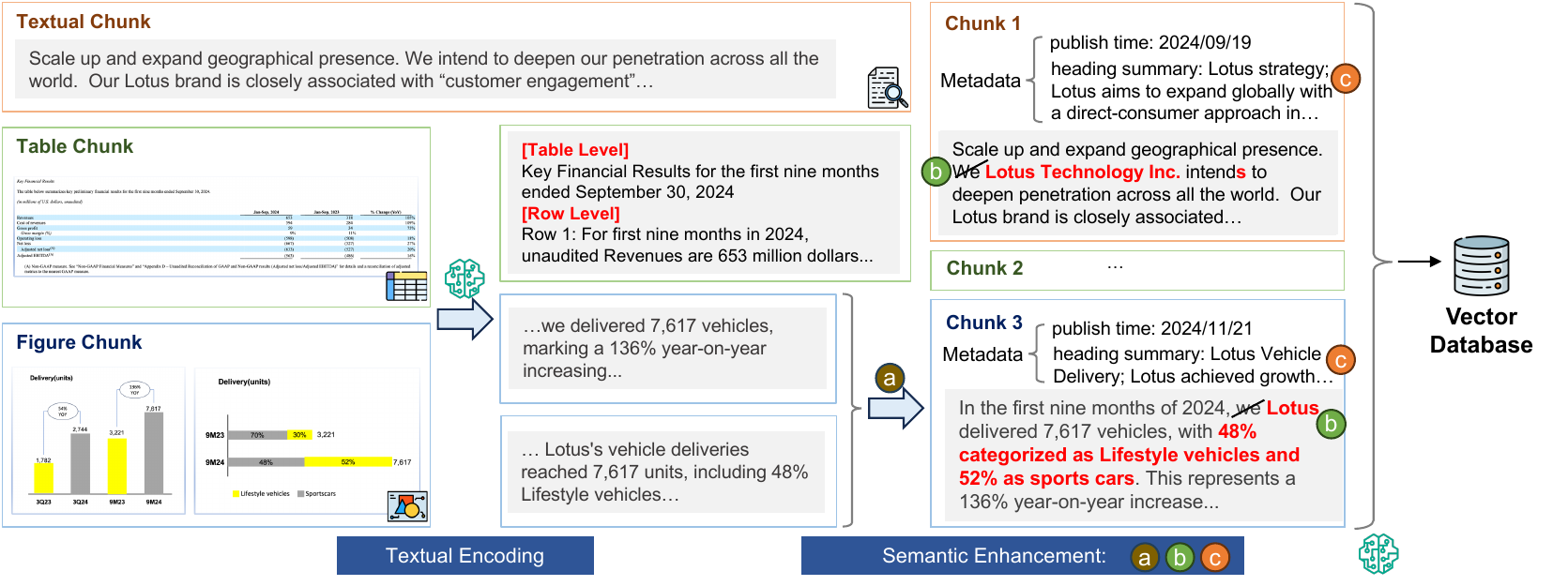} \caption{The \ourmodel{} FFP pipeline. \textbf{Textual Encoding:} PDF parsing tools extract multi-modal chunks (text, figures, tables), which are then converted into textual representations using large language models (LLMs). \textbf{Semantic Enhancement:} The textual representations are further refined by (a) eliminating redundant chunks through pairwise similarity comparison, (b) resolving co-reference within subtitle sections, and (c) generating subtitle-section-based summaries as metadata. These enriched chunks are then embedded in a vector database for further processing.} 
    \label{fig:pre-processing} 
\end{figure*}

% \subsection{Background}
\vspace{-1mm}
\subsection{Financial Filings Pre-processing pipeline} \label{sub: ffp}
We create a Financial Filings Pre-processing~(FFP) pipeline to convert multi-modal financial filings into machine-readable text chunks. \autoref{fig:pre-processing} illustrates how FFP operates in two steps: textual encoding and semantic enhancement.
% Our pre-processing procedure, called the Financial Filings Pre-processing Pipeline (FFP), is a linear workflow designed to convert multi-modal financial filings including tables and figures %specifically those filed to the SEC, 
% 重点是SEC这件事情提一次就可以了
% into machine-understandable text chunks. \autoref{fig: pre-processing} illustrates the self-contained two-step pipeline, including textual encoding and semantic enhancement.
% 不太需要面面俱到hhh

Formally, let $\mathcal{C}$ denote the original set of multi-modal chunks extracted from a financial document, divided based on its chapters and a pre-defined token length. Each chunk $s_i = \left(t_i, m_i\right) \in \mathcal{C}$ consists of raw content $t_i$ and metadata $m_i$. %, which includes the page number, publication time, and section heading. 
The goal of the FFP is to transform chunks in $\mathcal{C}$ into $\tilde{\mathcal{C}}$, a self-contained and condensed format that is better suited for downstream information retrieval tasks.

% \begin{itemize}
%     \item $t$: Raw text content
%     \item $\mathbf{e}_d$: Dense embedding (e.g., from FAISS-indexed encoder)
%     \item $\mathbf{e}_s$: Sparse embedding (e.g., BM25 lexical representation)
%     \item $\mathbf{e}_{meta}$: Metadata dense embedding
%     \item $H={\mathbf{h}_1, ..., \mathbf{h}_k}$: Set of hypothetical answer embeddings
% \end{itemize}

\paragraph{Textual Encoding of Multi-Modal Documents.}
% First, the FFP pipeline parses multi-modal financial documents into machine-readable textual formats. Using open-source PDF parsing tools such as MinerU, each document is decomposed into a list of chunks in reading order, each of which is labeled with its corresponding type (i.e., text, figure, or table). Inspired by~\citep{JimenoYepes2024FinancialRC}, we further transform figure and table chunks within the list into narrative statements in textual format, enabling them to be processed by the LLM.
The FFP pipeline first converts multi-modal financial documents into structured, machine-readable textual formats. %Using open-source PDF parsing tools such as MinerU~\citep{2024mineru}, each document is decomposed into a sequence of chunks following the natural reading order. 
Following the natural reading order, we use open-source PDF parsing tools, such as MinerU~\citep{2024mineru}, to decompose each document into a sequence of chunks. Each chunk is labeled with its corresponding type (i.e., text, figure, or table). Inspired by~\citet{JimenoYepes2024FinancialRC}, we further transform figure and table chunks into descriptive textual narratives with large vision-language models. Specifically, figures are converted into structured captions summarizing their key insights, while tables are rewritten as textual statements capturing their primary data trends and relationships~\citep{zhang2023moqagpt}. This transformation enables large language models (LLMs) to effectively interpret and utilize multi-modal content.

\paragraph{Semantic Enhancements.}
To improve the quality and coherence of the extracted texts, the FFP pipeline applies the following three key semantic enhancement steps.
% (a) eliminating redundant chunks, (b) resolving anaphora within subtitle sections, and (c) generating section-based summaries as metadata.

\begin{enumerate}[leftmargin=*]
    \item \textbf{Redundant chunk de-duplication.} We first de-duplicate documents to reduce redundancy, ensuring that duplicate content does not limit the diversity of retrieved information. 
    For any two chunks $s_{i}, s_{j} \in \mathcal{C}$ such that $i \neq j$, we randomly discard $s_i$ or $s_{j}$ if the cosine similarity of the dense embeddings of their raw texts $\cos\left(\mathbf{d}_{i}, \mathbf{d}_{j}\right)$ exceeds a predefined threshold $\tau_{\text{dedup}}$. 
    %For any two chunks $s_{i}, s_{j} \in \mathcal{C}$ where $i \neq j$, we remove $s_{j}$ if $\cos(\mathbf{d}_{i}, \mathbf{d}_{j}) > \tau_{dedup}$, where $\tau_{dedup}$ is a predefined similarity threshold, and $\mathbf{d}_i$ is the dense embedding of the raw text in the $i$-th chunk $t_i$.
    \item \textbf{Co-reference resolution.} Next, we disambiguate the textual context of each chunk by iteratively resolving co-references. Within each subtitle section, pronouns of entity mentions are replaced with their corresponding antecedents identified from preceding chunks. Specifically, for each chunk containing pronouns, we employ an LLM as a co-reference resolver to perform entity mention replacements. The LLM receives $k$ previous chunks to identify proper antecedents, where $k$ is fixed.
    % Formally, using an anaphora resolution function $R$, a chunk $s_i$ is updated as $\tilde{s}_i = R(s_{k,i}, \{s_{k,j} \mid s_{k,j} \in S_{k}, j < i\})$, where $\tilde{s}_{k,i}$ represents the chunk after resolving co-references.
    \item \textbf{Metadata generation.} 
    Due to the token-length limitation, document sections may be separated into multiple chunks where content may no longer be self-contained. Thus, we ensure that chunks within a section retain sufficient context from others in the same section for retrieval. We employ an LLM to summarize each section of the financial filing and append the generated summary to the metadata of every chunk obtained from that section. This ensures that each chunk retains enough contextual information while maintaining a minimal context length, enhancing the efficiency and accuracy of document retrieval.
    % Defining a context integration function $F$, we obtain the enhanced chunk as $\hat{s}_{k,i} = F(\tilde{s}_{k,i}, \{s_{k,j} \mid s_{k,j} \in S_{k}, j < i\})$.
    % Additionally, we generate a heading and summary for each subtitle section using a summarization function $G$, such that $h_k, s_k^{*} = G(\{s_{k,i} \mid s_{k,i} \in S_k\})$.
\end{enumerate}
\vspace{-0.5mm}
After the aforementioned process, for each chunk $s_i$, dense and sparse embeddings $\mathbf{d}_i$ and $\mathbf{p}_i$ are generated from $\tilde{t}_i$ using a text encoder and BM25, respectively. Furthermore, a dense embedding for the metadata $\mathbf{e}_{i}$ is also generated by a text encoder. The chunk is then represented as a 5-tuple $(\tilde{t}_i, \tilde{m}_i, \mathbf{d}_i, \mathbf{p}_i, \mathbf{e}_i)$. These enhancements ensure that extracted chunks are self-contained, contextually coherent, and semantically enriched, optimizing their utility for downstream processing. Further details regarding the selection of the multi-modal textualization model and the dense embedding model are found in Appendices~\ref{appx:vlm} and \ref{appx:embedding_model}, respectively.

\subsection{Multi-path Retrieval pipeline}
\label{sub:multi-path-retrieval-pipeline}
% We next introduce a Multi-path Retrieval~(MPR) pipeline that improves performance through a multi-path retrieval strategy by enhancing information retrieval through use of both metadata and semantic data. It comprises of two key modules used for query paraphrasing and for retrieving relevant answers from diverse sources.
We introduce a Multi-path Retrieval (MPR) pipeline that enhances performance by leveraging both metadata and semantic information during the retrieval process. The pipeline consists of two primary modules: a Query Paraphrasing Module and a Retrieval Module for obtaining relevant answers from diverse sources.

% % Should be ``an LLM'', not ``a LLM''
% \paragraph{Query Paraphrasing Module}
% We employ an LLM agent to rewrite each input query using \ourmodel{}'s specialized vector store, which serves as a knowledge base containing all company-related information. This process follows three key steps: 1) each query is split into multiple self-contained sub-queries, ensuring they can be independently answered by mapping query keywords to the corresponding knowledge segments %: $f(q_u)=\{q^{(1)}, \cdots, \allowbreak 
%  %q^{(n)}\}$; 
% 2) each query has its co-references resolved, converting it into a standardized English format where all pronouns are replaced with their corresponding subject names; 3) relevant context from the conversation history is integrated into queries referring to previous ones, maintaining completeness and clarity.
% Formally, recalling $S$ as the set of all pre-processed segments, %let $\mathcal{C}$ represent the set of preprocessed chunks. 
% the system paraphrases every user query $q_{u}$ into multiple subqueries $\{q^{(1)}, ..., q^{(n)}\}$ by mapping query keywords to the corresponding knowledge segments and to create paraphrased subqueries $\{q^{(1)*}, ..., q^{(n)*}\}$ by undergoing standardized English formatting and conversational history co-referencing: $\{q^{(1)}, ..., q^{(n)}\} \overset{f}{\rightarrow}\{q^{(1)*}, ..., q^{(n)*}\}$.

\paragraph{Query Paraphrasing Module}
An LLM agent is employed to reformulate each input query using \ourmodel{}'s specialized vector store, which serves as a comprehensive knowledge base containing all company-related information. This paraphrasing process involves three key steps:
\begin{enumerate}[leftmargin=*]
    \item \textbf{Query Decomposition:} Each query is segmented into multiple self-contained sub-queries. This ensures that each subquery can be independently addressed by mapping its keywords to the corresponding segments in the knowledge base.
    \item \textbf{Co-reference Resolution:} Co-references within each query are resolved by converting the text into standardized English. In this process, all pronouns are replaced with their corresponding antecedents.
    \item \textbf{Context Integration:} For queries that reference prior conversation history, relevant contextual information is incorporated to maintain completeness and clarity.
\end{enumerate}
\vspace{-1mm}
\paragraph{Retrieval Module}
% Traditional retrieval methods rely either on sparse or dense retrieval, which are difficult to adapt to specific domain.
% Dense retrievers require fine-tuning with large amounts of resources while sparse retrievers are often heuristic-based, potentially resulting in faulty retrievals.
% To overcome these limitation, we add metadata and hypothesis-based retrieval, forming a multi-path retrieval system. Thus our MPR pipeline has four components:
Traditional retrieval methods typically rely on either sparse or dense retrieval techniques, each presenting challenges when applied to domain-specific tasks. Dense retrievers generally require extensive fine-tuning with substantial resources, whereas sparse retrievers often depend on heuristic rules that can lead to inaccuracies. To address these limitations, we incorporate both metadata-based and hypothesis-based retrieval strategies, resulting in a multi-path retrieval system. Consequently, our MPR pipeline comprises four components:
\begin{enumerate}[leftmargin=*]
    \item \textbf{BM25 Sparse Retriever:} This component employs the BM25 scoring function to compute relevance based on tf-idf between the query and documents~\citep{robertson2009probabilistic}.
    \item \textbf{BGE-M3 Dense Retriever:} Utilizing the BGE-M3 model~\citep{chen2024bge}, this retriever calculates the cosine similarity between query embeddings and chunk embeddings. The retrieval process is accelerated by FAISS~\citep{johnson2019billion}, ensuring high-performance search.
    \item \textbf{Metadata Retriever:} Leveraging the same BGE-M3 model, this module computes the cosine similarity between the query and metadata embeddings, facilitating the integration of structured metadata into the retrieval process.
    \item \textbf{HyDE Retriever:} This component computes the cosine similarity between hypothesis and chunk embeddings.
\end{enumerate}
We collect all results selected by those retrievers into a chunk set, which serves as the output of the retrieval module.
% Formally, for each subquery $q^{(i)}$, path $j$ of the multi-path retrieval system retrieves a candidate chunk set: $$\mathcal{R}^{(j)}q^{(i)}=\{s \in C|s \text{ retrieved by }j\text{-th path for }q^{(i)}\},$$ such that the %via independent retrieval functions and hence 
% the overall chunk sets for the query is the union of all subquery chunk sets: $\mathcal{R}^{(j)}{q}=\bigcup \mathcal{R}^{(j)}q^{(i)}$.
% 這個地方的縮進是可以去掉的，具体可以GPT一下

% Our metadata retrieval approach addresses high-level search needs, such as QA tasks requiring reasoning across multiple chunks. Traditional retrieval struggles to capture all relevant chunks, so we use metadata summaries under the same subtitle (generated via FFP) to enhance retrieval coherence.
% Our metadata retrieval approach creatively concatenates each chunk's heading and summary to form a metadata embedding. In our framework, every chunk within a semantic segment is assigned an identical heading-summary representation, ensuring that the retrieval of a single metadata instance inherently retrieves all associated chunks. This design not only reinforces the contextual linkage among chunks from the same document segment but also effectively mitigates retrieval ambiguity when dealing with multiple documents. By grounding the metadata in key structural elements of the content, our method enhances the overall precision of the retrieval process, particularly in scenarios where traditional methods might suffer from document-level confusion.
Our metadata retrieval approach is designed to address high-level search tasks, such as question answering that requires reasoning across multiple text segments. Traditional retrieval methods often struggle to capture all relevant chunks; therefore, we incorporate metadata summaries (generated via FFP) under a consistent subtitle to enhance retrieval coherence.
Specifically, our approach concatenates each chunk's heading and summary to form a unified metadata embedding. In our framework, every chunk within a semantic segment is assigned an identical heading-summary representation, ensuring that the retrieval of a single metadata instance inherently retrieves all associated chunks. This design not only reinforces the contextual linkage among chunks from the same document segment but also effectively mitigates retrieval ambiguity when dealing with multiple documents.

The HyDE retriever is designed to further address recall issues. In certain scenarios, target chunks and user queries exhibit only weak correlations, even when the queries are optimally formulated. In such cases, the HyDE retriever leverages Hypothetical Document Embeddings\cite{zhang2024exploring}, a specialized query expansion technique which uses an LLM-generated response to the user query as a hypothesis for retrieving similar chunks.
To further enhance HyDE's performance, we augment the LLM's domain-specific knowledge through instruction fine-tuning on our custom Company dataset, following the InPars method~\citep{10.1145/3477495.3531863}. Additionally, a smaller model is subsequently trained on a synthesized dataset to emulate the generative capabilities of larger models. As a result, HyDE not only improves the matching accuracy between the generated hypothetical responses and the target chunks but also bolsters overall retrieval performance across diverse content types.

% \paragraph{Chunk Bundling Module} After the chunks are retrieved, the MPR pipeline introduces chunk expansion mechanism to
% enhance contextual coherence across sequential chunks.
% This is particularly crucial in financial documents, where key topics and discussions often extend beyond individual chunks or paragraphs.
% Initially, each chunk is treated as an independent unit. These units are iteratively expanded by evaluating adjacent chunks—preceding, following, or both—based on a pre-defined semantic similarity threshold.
% % This adaptive aggregation ensures that relevant contextual information is retained, improving the accuracy and consistency of downstream processing.
% Formally, for each initial chunk \( s_{i} \), a group \( g_{i} = \{s_{i}\} \) is created and iteratively expanded by incorporating adjacent chunks \( s_{i, \text{adj}} = \{s_{i-1}, s_{i+1}\} \) if \( \phi (s_{adj}, g^{(i)}) \geq \tau_{\text{exp}} \), where \(\phi (\cdot)\) is the sigmoid function and \( \tau_{\text{exp}} \in [0,1] \) is an empirically determined threshold. To maintain contextual relevance and prevent excessive aggregation, a maximum limit of five chunks per group is enforced.
\paragraph{Chunk Bundling Module} 
After retrieval, the MPR pipeline employs a chunk expansion mechanism to enhance contextual coherence among sequential chunks. This approach is particularly critical in financial documents, where key topics and discussions often span multiple chunks or paragraphs. Initially, each chunk is treated as an independent unit. These units are then iteratively expanded by evaluating adjacent chunks (preceding, following, or both) based on a predefined semantic similarity threshold.
Formally, for each initial chunk \( s_{i} \), we combine it with adjacent chunks $s_{i-1}$ or $s_{i+1}$ if the cosine similarity of their dense embeddings, namely $\cos(\mathbf{d}_i, \mathbf{d}_{i-1})$ or $\cos(\mathbf{d}_i, \mathbf{d}_{i-1})$) exceeds an empirically determined threshold $\tau_{\text{exp}}$.
% The group is subsequently expanded by incorporating adjacent chunks \( s_{i,\text{adj}} = \{s_{i-1}, s_{i+1}\} \) if the similarity measure \( \phi(s_{\text{adj}}, g^{(i)}) \geq \tau_{\text{exp}} \), where \( \phi(\cdot) \) denotes the sigmoid function and \( \tau_{\text{exp}} \in [0,1] \) is an empirically determined threshold. To preserve contextual relevance and avoid excessive aggregation, a maximum limit of five chunks per group is enforced.

\subsection{Document Re-ranking}
\label{sub: reranker}
The Document Re-ranker (DRR) in \ourmodel{} acts as an intelligent filter, addressing shortcomings of initial retrieval to enhance accuracy, coherence, and reliability in final generated responses. We employ the \texttt{BAAI/bge-reranker-v2-Gemma} model based on its superior performance in both English and multilingual applications. 
% Leveraging a cross-encoding architecture, it outperforms traditional dual-tower models in semantic comprehension. 
Within \ourmodel{}, the re-ranker refines retrieved documents before passing them to the LLM, providing more relevant contextual input for answer generation. Specifically, the Document Re-ranker scores candidate chunks $s\in \mathcal{R}_{q}$ as $\mathcal{K}\left(s, q\right)$ using a cross-encoder:
\begin{equation}
    \mathcal{K}\left(s_i, q\right) = \sigma\left(\mathbf{w}^\top \text{CrossEncoder}\left(\text{concat}\left(q, \tilde{t}_i\right)\right)+ \beta f\left(\tilde{m}_i\right)+b\right),
\end{equation}
where \(\textbf{w}^{\top} \text{CrossEncoder}(\cdot)+b\) is a Transformer-based re-ranker scoring contextual alignment~\citep{déjean2024thoroughcomparisoncrossencodersllms, askari2023injectingbm25scoretext}, and $f\left(\tilde{m}_i\right)$ processes metadata to output a scalar bonus, scaled by $\beta$. The final score is transformed using a sigmoid function $\sigma$.

\paragraph{Time bonus} When ranking chunks for relevance, a common challenge arises when two similar chunks lead to favoring older content instead of the new, potentially more desirable entries. Recognizing that metadata, particularly publication dates, is crucial for evaluating chunk relevance, we incorporate a time bonus for chunk $s_i$, $f\left(\tilde{m}_i\right)$, into the overall ranking score in DRR. Specifically, if the publication date of the chunk is within a year of the query time, there will be an additional bonus. The closer the publication date is to the query time, the larger this bonus will be. This mitigates the issue where the DRR favors older contents over new one when the two chunks have identical embeddings. 

%In ranking chunks for relevance, a common challenge arises when two similar chunks lead to favor over older content instead of the new, potentially more desirable entries. Recognizing that metadata, particularly publication dates, is crucial for evaluating chunk relevance, we incorporate a time bonus for chunk $s$, $f\left(m^{(s)}\right)$, into the overall ranking score in DRR. Our approach employs a simple method that compares each chunk’s publication date with the query timestamp. Chunks with publication dates closest to the query time receive an additional bonus, effectively prioritizing more current and relevant entries.

\paragraph{Direct Preference Optimization}
To further enhance ranking performance, we fine-tune the DRR using Direct Preference Optimization (DPO). This method frames preference learning as a binary classification task, where training data consists of preference pairs: a preferred (positive) response $s^{+}$ and a rejected (negative) response 
$s^{-}$. The model learns to prioritize high-quality responses by minimizing the cross-entropy loss:
\begin{equation}
\mathcal{L}{(\text{DPO})} = -\mathbb{E}{(q, s^+, s^-)}\left[\log \frac{\mathcal{K}(s^+, q)}{\mathcal{K}(s^+, q) + \mathcal{K}(s^-, q)}\right],
\end{equation}
ensuring that our re-ranker effectively distinguishes and prioritizes the most relevant document chunks.

The DRR pipeline is iteratively implemented as follows: \begin{enumerate}[leftmargin=*]
    \item \textbf{Retrieval and Annotation}: Extensive results are retrieved using FAISS in the Multi-Path Retrieval pipeline (Section~\ref{sub:multi-path-retrieval-pipeline}) for a rich set of candidate documents in each query. Retrieved documents are annotated to create preference samples for DPO, distinguishing between relevant and irrelevant matches. The annotated pairs are then organized into a structured dataset with the balance between positive and negative samples maintained to prevent training bias.
    \item \textbf{Evaluation}: Multiple evaluation metrics, including Normalized Discounted Cumulative Gain (NDCG) and Mean Reciprocal Rank (MRR)~\citep{schwartz2021ensemblemrrndcgmodels}, are employed to verify the model's ability to determine document relevancy and hence improve DRR's performance.
    \item \textbf{Model Adjustment}: When the model fails to perform on a new test set, the retrieved documents are re-annotated to better represent the underlying patterns and relationships via a careful review and adjustment of training data. The model is retrained on top of the previous version using the updated annotations, refining the ranking ability. This is repeated until a satisfactory performance on the new test set is attained.
\end{enumerate}

After document re-ranking, the RAG pipeline in \ourmodel{} follows the traditional method to handle user queries, retrieve related data from the vector database, and generate answers accordingly~(\autoref{fig:overview}). 

\section{Experiments}
We conduct experiments across three settings to evaluate \ourmodel{}: (1) assessing the Multi-path Retrieval pipeline, (2) evaluating the Document Re-ranker, and (3) measuring overall answer quality.
% For all experiments, we utilize FinanceBench~\citep{islam2023financebench}, a standardized benchmark in financial question answering, along with a manually annotated in-depth dataset to demonstrate the effectiveness of \ourmodel{}.

\subsection{Datasets and Baseline}

To evaluate performance, a manually annotated dataset is typically regarded as the ``golden dataset,'' with high-quality entries serving as the ground truth~\citep{kluai}. We utilize two such datasets:

\paragraph{FinanceBench}
FinanceBench~\citep{islam2023financebench} is a financial question-answering dataset created to evaluate the performance of LLMs on open-book tasks. It includes 150 open-source questions relating to traded companies and provides raw financial documents in PDF format containing straightforward answers. We use 149 questions from FinanceBench and construct a single vector store by processing 83 documents that contain correct answers for these questions with FFP. A fixed length of 256 tokens is adopted when separating semantic segments into chunks. We compare our proposed RAG system with three baseline methods, namely \citet{islam2023financebench}, \citet{JimenoYepes2024FinancialRC} and~\citet{setty2024improvingretrievalragbased} with their best-matching variants \textit{Shared Vector Store}, \textit{Base256}, and \textit{Reranker}, respectively. 
%We choose the baseline configurations as (\textit{Reranker} ~\citep{setty2024improvingretrievalragbased}, \textit{Base256} ~\citep{JimenoYepes2024FinancialRC}, and \textit{Shared Vector Store} ~\citep{islam2023financebench}) 
These variants generally follow a setup similar to ours, utilizing a single vector store for all filings in FinanceBench, a chunk size of 256 tokens, as well as the retriever before generating answers. The baseline \textit{reranker} is combined with a cross-encoder reranker which best aligns with our configuration.

\paragraph{In-depth Single Company Dataset}
% Unlike FinanceBench, this custom question-answering dataset focuses on 12 filings from a single company. We chunk these files into a single vector database with a fixed size of 200 tokens in one chunk. 
% We manually collect 75 question-answer pairs from these SEC 6-K and F-1 filings, along with an average of 10 relevant text chunks per question that contain the most pertinent passages. The questions span various aspects of Lotus Technology Inc., including products, sales data, history, and corporate strategy. We then engage with domain experts to carefully review the manually annotated chunks, so that the answers can be considered as the ground truth.  The complexity of these questions often requires gathering information across various documents or document sections, making it challenging to locate answers within a single chunk.
Apart from FinanceBench, we construct a custom QA dataset using 12 filings from a single company. These filings are segmented into a vector database, with each chunk fixed at 200 tokens.
We manually compile 75 question-answer pairs from SEC 6-K and F-1 filings, associating each question with an average of 8.7 relevant text chunks containing the most pertinent passages. The questions cover various aspects of Lotus Technology Inc., including products, sales data, history, and corporate strategy. To ensure accuracy, we engage domain experts to review the manually annotated chunks, establishing them as the ground truth. The complexity of these questions often necessitates aggregating information from multiple documents or sections, making it challenging to locate answers within a single chunk. We refer to this dataset henceforth as the Company dataset.

% \paragraph{Testing Set Setup}
% To construct our testing set, we compile a comprehensive collection of 75 test questions spanning various aspects of our domain knowledge. For each question, we manually identify and annotate an average of 10 relevant text chunks that contain the most pertinent passages, ensuring that they contain the necessary information for the LLM to generate complete and accurate responses.

\subsection{Experiment Settings}

\subsubsection{MPR settings}
We present the experiment settings for the Multi-Path Retrieval (MPR) pipeline, following the order of its retrieval workflow. After the query process, \ourmodel{} selects the top-$K$ results from the vector database to match the query.

In our HyDE retriever, we train a \texttt{Qwen‑2.5‑7B‑Instruct} to follow the HyDE instructions and acquire finance knowledge from the 72B model. To synthesize the training dataset, we first generate question-document pairs using \texttt{Qwen-2.5-72B-Instruct} and select the top 2,000 pairs based on similarity scores. Two hypothetical chunks are generated for each pair, encapsulating essential domain-specific nuances. After training, the model produces three hypothetical answers for each question during testing, which are then used to retrieve the corresponding set of top-$K$ chunks.

For the metadata retriever, the process selects the top-$K$ metadata entries and includes all chunks from the corresponding semantic segments as candidates. Accordingly, to maintain a consistent number of retrieved chunks across different methods, we adopt varying top‑K values. 

Finally, unique chunks are collected across different retrieval methods, after which we select $\tau_{\text{exp}}=0.85$ for the chunk expansion threshold with a limit of $\mathcal{K}=5$, to include the neighbor chunk with sufficient similarity.

\subsubsection{DRR settings}
We assess the impact of Document Re-ranker on retrieval effectiveness using different configurations (Top $K$ + Different Re-ranker), as detailed in \autoref{fig:ranking_comparison}. Here, Top $K$ denotes the number of top-reranked chunks by the re-ranker. For re-ranking, we compare our \textbf{fine-tuned DRR}, trained on single-company financial filings, against \texttt{BAAI/bge-reranker-v2-gemma}, a general re-ranker. Our model is fine-tuned using Direct Preference Optimization (DPO) with a contrastive learning objective. The training dataset comprises annotated query-document pairs, each containing one relevant chunk and multiple irrelevant ones. This enables the model to adapt to SEC filings' terminology and structure, ranking relevant chunks higher for improved retrieval precision. The experiment aims to evaluate whether a domain-specific re-ranker enhances retrieval effectiveness.
\subsubsection{Response settings}
We evaluate the end-to-end question answering (QA) performance of \ourmodel{} by comparing it against question answering approach proposed in ~\citep{setty2024improvingretrievalragbased, JimenoYepes2024FinancialRC, islam2023financebench} on the FinanceBench dataset. Additionally, we conduct QA accuracy evaluation on our in-depth company dataset. 
For FinanceBench, we construct a single vector store (one store for all filings, chunk size = 256) by processing financial PDF documents using the Financial Filings Preprocessing Pipeline (FFP) and apply our multi-path retriever combined with a fine-tuned document re-ranker to retrieve relevant chunks before generating answers. This setup on FinanceBench resembles the setup of \textit{Reranker}~\citep{setty2024improvingretrievalragbased}, \textit{Base256}~\citep{JimenoYepes2024FinancialRC}, and \textit{Shared Vector Store}~\citep{islam2023financebench}. 
In addition, we conduct a separate end-to-end evaluation on the Company dataset, where we measure QA accuracy by comparing generated answers against ground truth answers. For response generation, we set $K=5$ for each sub-query to efficiently manage LLM token consumption. This threshold is justified by empirical evidence indicating that retrieving five chunks is generally sufficient to generate effective answers~\citep{liu2023lostmiddlelanguagemodels}.

\begin{table*}[ht!]
    \centering
    \caption{Retrieval Performance on the company dataset. ``Avg Num Retrieved" represents the average number of chunks in the retrieved set. Recall, Precision, and F1 scores are calculated based on the relevant chunks retrieved. The bold denotes the best Recall, Precision and F1 produced from \ourmodel{}.}
    \label{tab:retrieval_ablation}
    \resizebox{\textwidth}{!}{ % Ensures table fits the page width
    \begin{tabular}{p{3.4cm} p{5.5cm} cccc} % Adjusted column widths
        \toprule
        \small \textbf{Category} & \textbf{Method} & \textbf{Avg Recall} & \textbf{Avg Precision} & \textbf{Avg F1} & \textbf{Avg Num Retrieved} \\
        \midrule
        \multirow{2}{*}{\small \textbf{FAISS Retrieval}} 
        & FAISS (Baseline) & 0.8194 & 0.0941 & 0.1646 & 75.6 \\
        & \quad + Bundle Expansion (Exp) & 0.8103 & 0.0999 & 0.1733 & 70.25 \\
        \midrule
        \multirow{1}{*}{\small \textbf{BM25 Retrieval}}
        & \quad + BM25 & 0.8452 & 0.1147 & 0.1949 & 66.24 \\
        \midrule
        \multirow{1}{*}{\small \textbf{Metadata Retrieval}} 
        & \quad + Metadata & 0.8573 & 0.1198 & 0.2037 & 64.87 \\
        % & \quad + Metadata + BM25 & 0.9113 & 0.1204 & 0.2052 & 69.91 \\
        \midrule
        \multirow{3}{*}{\small \textbf{HyDE Retrieval}} 
        & \quad HyDE-1: HyDE(Qwen7B) & 0.8228 & 0.1076 & 0.1830 & 69.09 \\
        & \quad HyDE-2: HyDE(Qwen7B-SFT) & 0.8567 & 0.1072 & 0.1831 & 72.92 \\
        & \quad HyDE-3: HyDE(Qwen72B) & 0.8323 & 0.1078 & 0.1844 & 69.77 \\
        \midrule
        \small \textbf{\ourmodel{}} & \quad + BM25+Metadata+HyDE-2 + Exp & \textbf{0.9251} & \textbf{0.1272} & \textbf{0.2156} & 68.75 \\
        \bottomrule
    \end{tabular}
    }
    \vspace{-0.5em}
\end{table*}

\vspace{-2mm}
\subsection{Evaluation}
\paragraph{MPR Evaluation} Our MPR pipeline is evaluated using various combinations of retriever paths by analyzing document chunk recommendation metrics (Recall, Precision, F1) on the in-depth dataset. By introducing additional retriever paths, we expect improvements in recall rates due to a more diverse information coverage~\citep{zhu2024rageval}.
\begin{figure}[t!]
  \centering
  \includegraphics[width=0.58\linewidth]{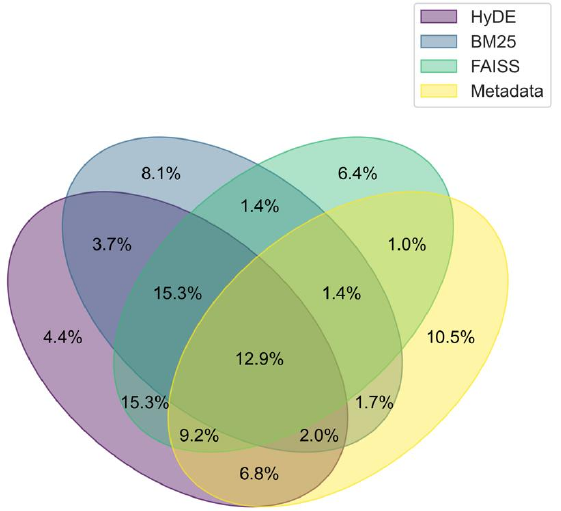}
  \caption{The percentage of relevant chunks selected by retrieval methods relative to the total number of retrieved relevant chunks. %The numbers in the overlapping regions indicate the percentage of relevant chunks retrieved by two or more retrieval methods.
  }
  \Description{Overlap of Positive Chunks across retrieving methods.}
  \label{fig:overlap}
  \vspace{-0.4cm}
\end{figure}
\paragraph{DRR Evaluation}
We assess the effectiveness of the Document Re-ranker by comparing the retrieved chunks with the manually annotated ground truth dataset. The re-ranker prioritizes and selects up to $K$ chunks per sub-query, where $K$ is an adjustable parameter. We then measure the re-ranking performance with Precision, Normalized Recall, Mean Reciprocal Rank (MRR), and Binary Normalized Discounted Cumulative Gain (nDCG). To ensure a fair assessment, we introduce an adjusted \textbf{Normalized Recall} metric:
\begin{equation}
    \text{Normalized Recall}=\frac{ \text{ correctly retrieved chunks}}{\min(  \text{ground truth chunks}, K)} \; \nonumber.
\end{equation}
We cap denominator to $\min(\text{ground truth chunks}, K)$, making it an upper bound for recall calculation. Traditional recall would underestimate and unfairly penalize the model if we were to divide by the actual number of ground truth chunks, despite the reranker functioning optimally within its retrieval limit.

\paragraph{Response Evaluation}
We compare between the quality of system-generated responses and expert-provided answers through automatic evaluation using \texttt{GPT-4o}. Specifically, we use the \textit{CorrectnessEvaluator} module from LlamaIndex, which utilizes \texttt{GPT-4o} to systematically score generated responses based on alignment of FinSage's outputs with ground truth reference answers. This follows prior work~\citep{DBLP:journals/corr/abs-2307-08161, naismith-etal-2023-automated, zheng2023judgingllmasajudgemtbenchchatbot} which utilize language models to judge the quality of system-generated responses. To complement automatic evaluation, our team manually labeled the responses after careful discussion. Human evaluation focuses on factual correctness, coherence, and completeness, however, minor factual inconsistencies may be tolerated if they do not alter the overall meaning or intent of the response. Details are provided in the appendix. These two metrics together provide a measurement to assess the \ourmodel{}'s ability to generate responses that align with the annotated ground truth answers both textually and semantically.

\begin{figure*}
    \includegraphics[width=\linewidth]{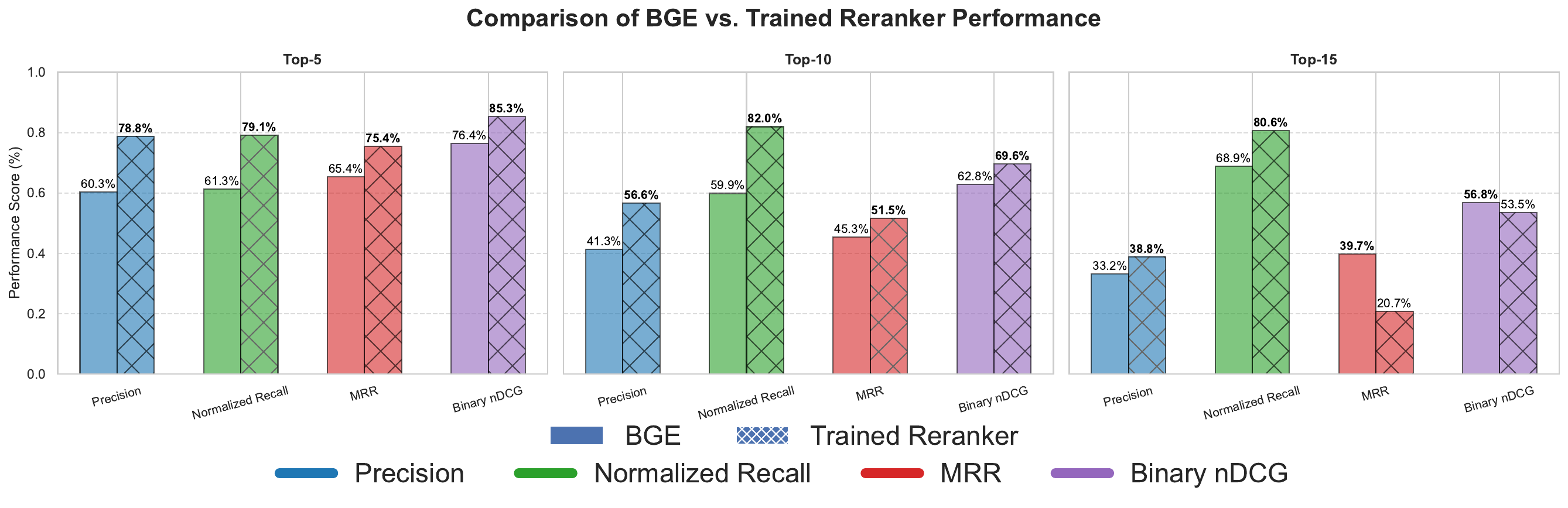} \vspace{-8mm} \caption{\small Performance of different re-rankers under different retrieval settings. }
    \label{fig:ranking_comparison}
\end{figure*}
\vspace{-1.5mm}
\section{Results and Analysis}
% We conduct experiments on the multi-path Retrieval Pipeline and Document Re-ranker to show the effectiveness of each crucial component in the \ourmodel{}. We also present the  Our evaluation is performed on the in-depth single company dataset, bringing more challenges to identifying positive document chunks for the query.

% \begin{figure*}[h!]
%   \centering
%   \includegraphics[width=\textwidth]{image/recall_comparison.pdf}
%   \caption{TODO: Comparison between different multi-path retrievers. Baseline uses only one query to retrieve 40 chunks and hyde uses an additional three hypothetical answers as queries and retrieves 10 chunks each. Faissonly uses dense retriever only and multi includes other retrieving methods like sparse BM25 and metadata retrieving. Expland means the bundling is enabled and chunks can be added from its neighbor chunks. Full means retrieving 40 chunks for each query, which retrieves the most chunks among different approach.}
%   \label{fig:diagram}
% \end{figure*}

\subsection{MPR Results}
% Table~\ref{tab:retrieval_ablation} shows the retrieval performance of the MPR pipeline with multiple paths, demonstrating gradual improvement in retrieval performance with an increasing number of paths. In the beginning, the FAISS-only retriever (the base case that only retrieves dense embeddings of chunks) achieves moderate performance of 81.94\% recall. The addition of BM25 to retrieve sparse embeddings elevates the recall score to 84.52\%. On the other hand, incorporating the metadata data retriever further increases the recall score to 85.73\%. Finally, \ourmodel{}, combining BM25, metadata retrieval, HyDE with \texttt{Qwen2.5-7B-SFT}, and bundle expansion achieves the best performance across all metrics and retrieves 92.51\% of all positive chunks. 

% Analyzing the detailed HyDE retrieval paths in Table \ref{tab:retrieval_ablation}, we find that our fine-tuned \texttt{Qwen2.5-7B-SFT} model surpasses the significantly larger \texttt{Qwen2.5-72B} base model with an outperforming recall rate, while still managing to achieve highly comparable precision and F1 score. This highlights the critical role of fine-tuning with high-quality domain-specific data, reinforcing the necessity of \ourmodel{} for precise and accurate question-answering in the complex structure of financial filings. Moreover, this resonates with the discussion where smaller fine-tuned language models are capable of performing as well as LLMs in specific domains~\citep{bucher2024finetunedsmallllmsstill}. 

\autoref{tab:retrieval_ablation} demonstrates \ourmodel{}'s superior performance through its novel multi-path retrieval architecture. While traditional single-path approaches like FAISS and BM25 show decent performance in isolation, they each capture only limited aspects of the complex financial domain based on superficial semantic matching. \ourmodel{}'s innovative combination of multiple specialized retrievers, particularly its integration of metadata-aware retrieval with fine-tuned HyDE hypothetical document generation, allows it to comprehensively capture both the structural and semantic aspects of financial documents. This synergistic approach enables \ourmodel{} to maintain high precision while significantly improving recall.

Our experimental results reveal the fine-tuned \texttt{Qwen2.5-7B-SFT} model surpasses a significantly larger \texttt{Qwen2.5-72B} counterpart across key metrics, particularly in recall performance. This highlights the critical role of fine-tuning with high-quality domain-specific data, reinforcing the necessity of \ourmodel{} for precise and accurate question-answering in the complex structure of financial filings. The strong performance of our smaller fine-tuned model aligns with emerging research demonstrating that well-adapted smaller language models can match or exceed the performance of larger models in specialized domains~\citep{bucher2024finetunedsmallllmsstill}, while offering substantial computational efficiency advantages.

%, while our fine-tuned model gets a competitive recall rate with the least number of total recall chunks, the model achieves a 15\% and 12\% improvement in the precision and f1 score respectively. \textcolor{red}{Add one line of analysis when confirm the fiure}

To better understand the uniqueness of each retrieval method, we analyze their overlap on the Company dataset. As shown in \autoref{fig:overlap}, while each method uniquely contributes to retrieving positive chunks, there is significant overlap, suggesting that hybrid approaches could leverage their combined strengths. Notably, FAISS and HyDE retrieve highly similar positive chunks, with an overlap rate of $52.7\%$ $(15.3\%+12.9\%+15.3\%+9.2\%)$, compared to their individual recovery rates of $62.9\%$ and $69.6\%$. This overlap likely arises because HyDE generates hypothetical documents by referencing the original document, preserving much of its context.
\vspace{-2mm}
\subsection{DRR Results}
% Figure~\ref{fig:ranking_comparison} shows that our Document Re-ranker significantly and consistently outperforms the general re-ranker across all settings in all metrics in top-5 and top-10 settings. Notably, our trained re-ranker retrieves relevant chunks illustrated by approximately $15\%$ improvement in recall for all setups. It is notable that the outperformance is not only on retrieving relevant chunks but also on mitigating irrelevant noises from being selected, as the precision shows an improvement from 5.6\% to 18.5\%. 

We compare re-rankers' performance in Figure~\ref{fig:ranking_comparison}, where we compare our trained re-ranker with \texttt{bge-reranker-v2-Gemma}. These results demonstrate that our Document Re-ranker significantly and consistently outperforms the general re-ranker across all settings and metrics in top-5 and top-10 configurations. Notably, our trained re-ranker retrieves more relevant chunks, evidenced by an approximate 15\% improvement in recall across all setups. Moreover, performance gains are not limited to retrieving relevant chunks, it also effectively filters out irrelevant information, leading to a substantial precision boost from 5.6\% to 18.5\%.

% It is worth noting that the precision, MRR, and binary nDCG all degrade when more chunks are extracted from the reranker, which shows a decreasing trend from 78.8\% to 38.8\%, 75.4\% to 20.7\% and 85.3\% to 53.5\%. This trend is consistent in the benchmark scenario. This shows that extracting more chunks increases the burden of the reranker, and would lead to a greater variance of chunk selections. This greater variance implies more irrelevant chunks to be extracted, i.e. drop of precision, and likely leads to a less optimal ranking, i.e. drop in MRR and Binary nDCG. The rapid degradation in the three metrics from Top-10 to Top-15 is consistent with studies showing that 5 - 10 chunks are generally suffice for effective answers~\citep{liu2023lostmiddlelanguagemodels}.

It is worth noting that precision, MRR, and binary nDCG all decline as more chunks are extracted by the reranker: precision drops from 78.8\% to 38.8\%, MRR from 75.4\% to 20.7\%, and binary nDCG from 85.3\% to 53.5\%. This consistent trend across benchmark scenarios indicates that extracting additional chunks increases the burden on the re-ranker, resulting in less optimal ranking in chunk selection during re-ranking.  Moreover, the rapid degradation in these metrics between Top-10 and Top-15 aligns with previous studies suggesting that retrieving only 5 to 10 chunks is generally sufficient for producing effective answers~\citep{liu2023lostmiddlelanguagemodels}.

% Overall, these findings emphasize the value of domain-specific fine-tuning for re-rankers in specialized retrieval tasks. While increasing the number of re-ranker extracted chunks $K$ enhances recall, it also demands a robust re-ranker to preserve ranking quality and minimize irrelevant retrievals. However, in the context of financial filings question-answering, recall should take precedence over precision to ensure that all queries receive comprehensive and informative responses.

\subsection{Response Results}

\begin{table}[h!]
    \centering
    \caption{Performance of End-To-End QA. The bold is from our implementation, which has the top LLM and Manual scores. Results with $*$ are from our experiments, while others are from the original papers. }
    \label{tab:response_fb}
    \begin{tabular}{p{1.6cm} p{3.6cm} cc} 
        \toprule
        \textbf{Dataset} & \textbf{Method} & \textbf{LLM} & \textbf{Manual}  \\
        \midrule
        \small{FinanceBench} & \citet{islam2023financebench}
        %~\textit{(Shared Vector Store)} 
        & \textbf{-} & \small{0.1900}   \\
        \small{FinanceBench} & \citet{JimenoYepes2024FinancialRC}%~\textit{(Base256)}
        & \small{0.3262} & \small{0.3688}  \\
        \small{FinanceBench} & \citet{setty2024improvingretrievalragbased}%~\textit{(Reranker)}
        & \small{0.2560} & \small{-} \\
        \small{FinanceBench} & \small{\ourmodel{}}$^{*}$ & \textbf{\small{0.4966}} & \textbf{\small{0.5705}} \\
        \midrule
        \small{Company} & \small{\ourmodel{}}$^{*}$ & \small{0.8533} & \small{0.8800}  \\
        \bottomrule
    \end{tabular}
    \vspace{-0.3cm}
\end{table}

\autoref{tab:response_fb} shows performance on the FinanceBench and Company datasets. \ourmodel{} achieves a significant improvement over prior methods on FinanceBench, with an LLM accuracy of 49.66\%, outperforming \citet{JimenoYepes2024FinancialRC} (32.62\%) and \citet{setty2024improvingretrievalragbased} (25.60\%, aligns most closely with our configuration). Similarly, manual evaluation also reflects an accuracy of 57.05\%, substantially higher than the closest competitor ~\citet{JimenoYepes2024FinancialRC} (36.88\%). These results indicate that FinSage’s retrieval and ranking mechanisms contribute significantly to improved answer quality.
In the Company dataset, \ourmodel{} also demonstrates strong performance, which achieves an LLM accuracy of 85.33\% and a manual accuracy of 88.00\%. The consistently higher scores across both datasets indicate that \ourmodel{} is particularly adept at processing structured financial filings with clear contextual information. By retrieving semantically relevant passages across multiple documents, \ourmodel{} significantly improves response accuracy, making it highly effective for complex financial question-answering tasks.

\begin{table}
    \centering
    \caption{Response Time Comparison (in seconds)}
    \label{tab:graph_light_rag}
    \resizebox{0.4\textwidth}{!}{
    \begin{tabular}{p{3cm} cccc}
        \toprule
        \textbf{Method} & \textbf{Mean} & \textbf{Median} & \textbf{Min} & \textbf{Max} \\
        \midrule
        GraphRAG & 16.90 & 15.27 & 9.86 & 40.67 \\
        LightRAG & 12.16 & 8.84 & 3.41 & 859.51 \\
        FinSage & 19.34 & 18.57 & 8.57 & 40.02 \\
        \bottomrule
    \end{tabular}
    }
    \vspace{-0.2cm}
\end{table}

\begin{table}
    \centering
    \caption{Faithful Evaluation Score Comparison}
    \label{tab:graph_light_rag_faithful_score}
    \resizebox{0.48\textwidth}{!}{
    \begin{tabular}{p{1.5cm} cccccc}
        \toprule
        \textbf{Method} & \textbf{Questions} & \textbf{Mean} & \textbf{Median} & \textbf{Min} & \textbf{Max} & \textbf{Pass \%} \\
        \midrule
        GraphRAG & 71(w/ 4 failures) & 3.46 & 3.50 & 2.0 & 5.0 & 42.50\% \\
        LightRAG & 75 & 2.45 & 2.00 & 1.0 & 5.0 & 13.67\% \\
        FinSage & 75 & 4.31 & 5.00 & 1.00 & 5.00 & 82.67\% \\
        \bottomrule
    \end{tabular}
    }
    \vspace{-0.5cm}
\end{table}

\subsection{Comparison with Other RAG Solutions}
We also compare our method with graph-based RAG solutions \textbf{GraphRAG} and \textbf{LightRAG} due to their mainstream usage. Table~\ref{tab:graph_light_rag} and \ref{tab:graph_light_rag_faithful_score} shows the advantages of Finsage in terms of response time and performance. This suggests that the additional complexity introduced by graph-based representation does not lead to improved response quality in our experimental context, which emphasizes the importance of balancing response speed with better performance when selecting RAG implementations for specific application domains. Detailed experimental settings and analysis can be found in Appendix~\ref{graphrag}.

% \subsection{Error Analysis}

\section{Conclusion}

% In this paper, we propose \ourmodel{}, a retrieval-augmented generation (RAG) solutions for question answering task on financial filings.
% % \ourmodel{} provides insights to automate process documents and ensure compliance to generate answsers in LLM. Experiments demonstrate that \ourmodel{} outperforms normal chunking techniques on financial filings, and has a production-ready document retrieving ability on multiple documents. In the future, we expect \ourmodel{} to further improve decision-making and agentic retrieval, providing greater control to the LLM to improve retrieval and generate more accurate answers.
% \ourmodel{} consists of a multi-modal data pre-processing pipeline, a domain aware multi-path retrieval system special designed for financial data, and a domain-specialized re-ranking module finetuned by DPO.
% Our system exhibited state-of-the-art performance on expert-curated financial filings question.
% We hope our system could help enterprises in solving regulatory compliance problem.
In this paper, we propose \ourmodel{}, a RAG solution for the question-answering task on financial filings. \ourmodel{} comprises a multi-modal data pre-processing pipeline, a domain-aware multi-path retrieval system specifically designed for financial data, and a domain-specialized re-ranking module fine-tuned using DPO. Our system demonstrated state-of-the-art performance on expert-curated questions derived from financial filings. Moreover, \ourmodel{} has been deployed as a financial question-answering agent in online meetings organized by affiliated entrepreneurs.

\section*{Acknowledgements}
This work is the result of a fruitful collaboration between \textbf{CG Matrix Technology Limited} and \textbf{SimpleWay AI}. We would like to express our sincere gratitude to both organizations for their generous support throughout this project. We are particularly grateful for the provision of essential computational resources, datasets, and invaluable domain expert assistance, which were crucial for the successful completion of this work. We also gratefully acknowledge Boyu Han (Standford University), Weien Li (MIT), and Zeyu Li (Nanyang Technological University) for their initial development of \ourmodel{}.

\bibliographystyle{ACM-Reference-Format}
\bibliography{sample-base}


\begin{thebibliography}{52}

%%% the Web address.

\ifx \showCODEN    \undefined \def \showCODEN     #1{\unskip}     \fi
\ifx \showISBNx    \undefined \def \showISBNx     #1{\unskip}     \fi
\ifx \showISBNxiii \undefined \def \showISBNxiii  #1{\unskip}     \fi
\ifx \showISSN     \undefined \def \showISSN      #1{\unskip}     \fi
\ifx \showLCCN     \undefined \def \showLCCN      #1{\unskip}     \fi
\ifx \shownote     \undefined \def \shownote      #1{#1}          \fi
\ifx \showarticletitle \undefined \def \showarticletitle #1{#1}   \fi
\ifx \showURL      \undefined \def \showURL       {\relax}        \fi
% invisible to TeX
\providecommand\bibfield[2]{#2}
\providecommand\bibinfo[2]{#2}
\providecommand\natexlab[1]{#1}
\providecommand\showeprint[2][]{arXiv:#2}

\bibitem[Agrawal et~al\mbox{.}(2024)]%
        {agrawal2024mindfulragstudypointsfailure}
\bibfield{author}{\bibinfo{person}{Garima Agrawal}, \bibinfo{person}{Tharindu Kumarage}, \bibinfo{person}{Zeyad Alghamdi}, {and} \bibinfo{person}{Huan Liu}.} \bibinfo{year}{2024}\natexlab{}.
\newblock \bibinfo{title}{Mindful-RAG: A Study of Points of Failure in Retrieval Augmented Generation}.
\newblock
\showeprint[arxiv]{2407.12216}~[cs.IR]
\urldef\tempurl%
\url{https://arxiv.org/abs/2407.12216}
\showURL{%
\tempurl}


\bibitem[Anonymous(2024)]%
        {anonymous2024tabmeta}
\bibfield{author}{\bibinfo{person}{Anonymous}.} \bibinfo{year}{2024}\natexlab{}.
\newblock \showarticletitle{TabMeta: Table Metadata Generation with {LLM}-Curated Dataset and {LLM}-Judges}. In \bibinfo{booktitle}{\emph{Submitted to ACL Rolling Review - June 2024}}.
\newblock
\urldef\tempurl%
\url{https://openreview.net/forum?id=NXYVm3AjG2}
\showURL{%
\tempurl}
\newblock
\shownote{under review}.


\bibitem[Asai et~al\mbox{.}(2024)]%
        {asai2024selfrag}
\bibfield{author}{\bibinfo{person}{Akari Asai}, \bibinfo{person}{Zeqiu Wu}, \bibinfo{person}{Yizhong Wang}, \bibinfo{person}{Avirup Sil}, {and} \bibinfo{person}{Hannaneh Hajishirzi}.} \bibinfo{year}{2024}\natexlab{}.
\newblock \showarticletitle{Self-RAG: Learning to retrieve, generate, and critique through self-reflection}. In \bibinfo{booktitle}{\emph{Proceedings of the Twelfth International Conference on Learning Representations}}.
\newblock


\bibitem[Askari et~al\mbox{.}(2023)]%
        {askari2023injectingbm25scoretext}
\bibfield{author}{\bibinfo{person}{Arian Askari}, \bibinfo{person}{Amin Abolghasemi}, \bibinfo{person}{Gabriella Pasi}, \bibinfo{person}{Wessel Kraaij}, {and} \bibinfo{person}{Suzan Verberne}.} \bibinfo{year}{2023}\natexlab{}.
\newblock \bibinfo{title}{Injecting the BM25 Score as Text Improves BERT-Based Re-rankers}.
\newblock
\showeprint[arxiv]{2301.09728}~[cs.IR]
\urldef\tempurl%
\url{https://arxiv.org/abs/2301.09728}
\showURL{%
\tempurl}


\bibitem[Barnett et~al\mbox{.}(2024)]%
        {barnett2024sevenfailurepointsengineering}
\bibfield{author}{\bibinfo{person}{Scott Barnett}, \bibinfo{person}{Stefanus Kurniawan}, \bibinfo{person}{Srikanth Thudumu}, \bibinfo{person}{Zach Brannelly}, {and} \bibinfo{person}{Mohamed Abdelrazek}.} \bibinfo{year}{2024}\natexlab{}.
\newblock \bibinfo{title}{Seven Failure Points When Engineering a Retrieval Augmented Generation System}.
\newblock
\showeprint[arxiv]{2401.05856}~[cs.SE]
\urldef\tempurl%
\url{https://arxiv.org/abs/2401.05856}
\showURL{%
\tempurl}


\bibitem[Besta et~al\mbox{.}(2024)]%
        {besta2024multiheadragsolvingmultiaspect}
\bibfield{author}{\bibinfo{person}{Maciej Besta}, \bibinfo{person}{Ales Kubicek}, \bibinfo{person}{Roman Niggli}, \bibinfo{person}{Robert Gerstenberger}, \bibinfo{person}{Lucas Weitzendorf}, \bibinfo{person}{Mingyuan Chi}, \bibinfo{person}{Patrick Iff}, \bibinfo{person}{Joanna Gajda}, \bibinfo{person}{Piotr Nyczyk}, \bibinfo{person}{Jürgen Müller}, \bibinfo{person}{Hubert Niewiadomski}, \bibinfo{person}{Marcin Chrapek}, \bibinfo{person}{Michał Podstawski}, {and} \bibinfo{person}{Torsten Hoefler}.} \bibinfo{year}{2024}\natexlab{}.
\newblock \bibinfo{title}{Multi-Head RAG: Solving Multi-Aspect Problems with LLMs}.
\newblock
\showeprint[arxiv]{2406.05085}~[cs.CL]
\urldef\tempurl%
\url{https://arxiv.org/abs/2406.05085}
\showURL{%
\tempurl}


\bibitem[Bonifacio et~al\mbox{.}(2022)]%
        {10.1145/3477495.3531863}
\bibfield{author}{\bibinfo{person}{Luiz Bonifacio}, \bibinfo{person}{Hugo Abonizio}, \bibinfo{person}{Marzieh Fadaee}, {and} \bibinfo{person}{Rodrigo Nogueira}.} \bibinfo{year}{2022}\natexlab{}.
\newblock \showarticletitle{InPars: Unsupervised Dataset Generation for Information Retrieval}. In \bibinfo{booktitle}{\emph{Proceedings of the 45th International ACM SIGIR Conference on Research and Development in Information Retrieval}} (Madrid, Spain) \emph{(\bibinfo{series}{SIGIR '22})}. \bibinfo{publisher}{Association for Computing Machinery}, \bibinfo{address}{New York, NY, USA}, \bibinfo{pages}{2387–2392}.
\newblock
\showISBNx{9781450387323}
\href{https://doi.org/10.1145/3477495.3531863}{doi:\nolinkurl{10.1145/3477495.3531863}}


\bibitem[Bucher and Martini(2024)]%
        {bucher2024finetunedsmallllmsstill}
\bibfield{author}{\bibinfo{person}{Martin Juan~José Bucher} {and} \bibinfo{person}{Marco Martini}.} \bibinfo{year}{2024}\natexlab{}.
\newblock \bibinfo{title}{Fine-Tuned 'Small' LLMs (Still) Significantly Outperform Zero-Shot Generative AI Models in Text Classification}.
\newblock
\showeprint[arxiv]{2406.08660}~[cs.CL]
\urldef\tempurl%
\url{https://arxiv.org/abs/2406.08660}
\showURL{%
\tempurl}


\bibitem[Cai et~al\mbox{.}(2022)]%
        {RAG2}
\bibfield{author}{\bibinfo{person}{Deng Cai}, \bibinfo{person}{Yan Wang}, \bibinfo{person}{Lemao Liu}, {and} \bibinfo{person}{Shuming Shi}.} \bibinfo{year}{2022}\natexlab{}.
\newblock \showarticletitle{Recent Advances in Retrieval-Augmented Text Generation}. In \bibinfo{booktitle}{\emph{Proceedings of the 45th International ACM SIGIR Conference on Research and Development in Information Retrieval}} (Madrid, Spain) \emph{(\bibinfo{series}{SIGIR '22})}. \bibinfo{publisher}{Association for Computing Machinery}, \bibinfo{address}{New York, NY, USA}, \bibinfo{pages}{3417–3419}.
\newblock
\showISBNx{9781450387323}
\href{https://doi.org/10.1145/3477495.3532682}{doi:\nolinkurl{10.1145/3477495.3532682}}


\bibitem[Chen et~al\mbox{.}(2024)]%
        {chen2024bge}
\bibfield{author}{\bibinfo{person}{Jianlv Chen}, \bibinfo{person}{Shitao Xiao}, \bibinfo{person}{Peitian Zhang}, \bibinfo{person}{Kun Luo}, \bibinfo{person}{Defu Lian}, {and} \bibinfo{person}{Zheng Liu}.} \bibinfo{year}{2024}\natexlab{}.
\newblock \showarticletitle{Bge m3-embedding: Multi-lingual, multi-functionality, multi-granularity text embeddings through self-knowledge distillation}.
\newblock \bibinfo{journal}{\emph{arXiv preprint arXiv:2402.03216}} (\bibinfo{year}{2024}).
\newblock


\bibitem[Chen et~al\mbox{.}(2022)]%
        {chen-etal-2022-murag}
\bibfield{author}{\bibinfo{person}{Wenhu Chen}, \bibinfo{person}{Hexiang Hu}, \bibinfo{person}{Xi Chen}, \bibinfo{person}{Pat Verga}, {and} \bibinfo{person}{William Cohen}.} \bibinfo{year}{2022}\natexlab{}.
\newblock \showarticletitle{{M}u{RAG}: Multimodal Retrieval-Augmented Generator for Open Question Answering over Images and Text}. In \bibinfo{booktitle}{\emph{Proceedings of the 2022 Conference on Empirical Methods in Natural Language Processing}}, \bibfield{editor}{\bibinfo{person}{Yoav Goldberg}, \bibinfo{person}{Zornitsa Kozareva}, {and} \bibinfo{person}{Yue Zhang}} (Eds.). \bibinfo{publisher}{Association for Computational Linguistics}, \bibinfo{address}{Abu Dhabi, United Arab Emirates}, \bibinfo{pages}{5558--5570}.
\newblock
\href{https://doi.org/10.18653/v1/2022.emnlp-main.375}{doi:\nolinkurl{10.18653/v1/2022.emnlp-main.375}}


\bibitem[Contributors(2024)]%
        {2024mineru}
\bibfield{author}{\bibinfo{person}{MinerU Contributors}.} \bibinfo{year}{2024}\natexlab{}.
\newblock \bibinfo{title}{MinerU: A One-stop, Open-source, High-quality Data Extraction Tool}.
\newblock \bibinfo{howpublished}{\url{https://github.com/opendatalab/MinerU}}.
\newblock


\bibitem[de~Souza P.~Moreira et~al\mbox{.}(2024)]%
        {moreira2024enhancingqatextretrieval}
\bibfield{author}{\bibinfo{person}{Gabriel de Souza P.~Moreira}, \bibinfo{person}{Ronay Ak}, \bibinfo{person}{Benedikt Schifferer}, \bibinfo{person}{Mengyao Xu}, \bibinfo{person}{Radek Osmulski}, {and} \bibinfo{person}{Even Oldridge}.} \bibinfo{year}{2024}\natexlab{}.
\newblock \bibinfo{title}{Enhancing Q\&A Text Retrieval with Ranking Models: Benchmarking, fine-tuning and deploying Rerankers for RAG}.
\newblock
\showeprint[arxiv]{2409.07691}~[cs.IR]
\urldef\tempurl%
\url{https://arxiv.org/abs/2409.07691}
\showURL{%
\tempurl}


\bibitem[Déjean et~al\mbox{.}(2024)]%
        {déjean2024thoroughcomparisoncrossencodersllms}
\bibfield{author}{\bibinfo{person}{Hervé Déjean}, \bibinfo{person}{Stéphane Clinchant}, {and} \bibinfo{person}{Thibault Formal}.} \bibinfo{year}{2024}\natexlab{}.
\newblock \bibinfo{title}{A Thorough Comparison of Cross-Encoders and LLMs for Reranking SPLADE}.
\newblock
\showeprint[arxiv]{2403.10407}~[cs.IR]
\urldef\tempurl%
\url{https://arxiv.org/abs/2403.10407}
\showURL{%
\tempurl}


\bibitem[Islam et~al\mbox{.}(2023)]%
        {islam2023financebench}
\bibfield{author}{\bibinfo{person}{Pranab Islam}, \bibinfo{person}{Anand Kannappan}, \bibinfo{person}{Douwe Kiela}, \bibinfo{person}{Rebecca Qian}, \bibinfo{person}{Nino Scherrer}, {and} \bibinfo{person}{Bertie Vidgen}.} \bibinfo{year}{2023}\natexlab{}.
\newblock \bibinfo{title}{FinanceBench: A New Benchmark for Financial Question Answering}.
\newblock
\showeprint[arxiv]{2311.11944}~[cs.CL]


\bibitem[Jacob et~al\mbox{.}(2024)]%
        {jacob2024drowningdocumentsconsequencesscaling}
\bibfield{author}{\bibinfo{person}{Mathew Jacob}, \bibinfo{person}{Erik Lindgren}, \bibinfo{person}{Matei Zaharia}, \bibinfo{person}{Michael Carbin}, \bibinfo{person}{Omar Khattab}, {and} \bibinfo{person}{Andrew Drozdov}.} \bibinfo{year}{2024}\natexlab{}.
\newblock \bibinfo{title}{Drowning in Documents: Consequences of Scaling Reranker Inference}.
\newblock
\showeprint[arxiv]{2411.11767}~[cs.IR]
\urldef\tempurl%
\url{https://arxiv.org/abs/2411.11767}
\showURL{%
\tempurl}


\bibitem[Jimeno-Yepes et~al\mbox{.}(2024)]%
        {JimenoYepes2024FinancialRC}
\bibfield{author}{\bibinfo{person}{Antonio Jimeno-Yepes}, \bibinfo{person}{Yao You}, \bibinfo{person}{Jan Milczek}, \bibinfo{person}{Sebastian Laverde}, {and} \bibinfo{person}{Renyu Li}.} \bibinfo{year}{2024}\natexlab{}.
\newblock \showarticletitle{Financial Report Chunking for Effective Retrieval Augmented Generation}.
\newblock \bibinfo{journal}{\emph{ArXiv}}  \bibinfo{volume}{abs/2402.05131} (\bibinfo{year}{2024}).
\newblock
\urldef\tempurl%
\url{https://api.semanticscholar.org/CorpusID:267547721}
\showURL{%
\tempurl}


\bibitem[Johnson et~al\mbox{.}(2019)]%
        {johnson2019billion}
\bibfield{author}{\bibinfo{person}{Jeff Johnson}, \bibinfo{person}{Matthijs Douze}, {and} \bibinfo{person}{Herv{\'e} J{\'e}gou}.} \bibinfo{year}{2019}\natexlab{}.
\newblock \showarticletitle{Billion-scale similarity search with GPUs}.
\newblock \bibinfo{journal}{\emph{IEEE Transactions on Big Data}} \bibinfo{volume}{7}, \bibinfo{number}{3} (\bibinfo{year}{2019}), \bibinfo{pages}{535--547}.
\newblock


\bibitem[Karpukhin et~al\mbox{.}(2020)]%
        {karpukhin2020dense}
\bibfield{author}{\bibinfo{person}{Vladimir Karpukhin}, \bibinfo{person}{Barlas Oguz}, \bibinfo{person}{Sewon Min}, \bibinfo{person}{Patrick Lewis}, \bibinfo{person}{Ledell Wu}, \bibinfo{person}{Sergey Edunov}, \bibinfo{person}{Danqi Chen}, {and} \bibinfo{person}{Wen-tau Yih}.} \bibinfo{year}{2020}\natexlab{}.
\newblock \showarticletitle{Dense Passage Retrieval for Open-Domain Question Answering}. In \bibinfo{booktitle}{\emph{Proceedings of the 2020 Conference on Empirical Methods in Natural Language Processing (EMNLP)}}. \bibinfo{pages}{6769--6781}.
\newblock


\bibitem[klu.ai({[n.\,d.]})]%
        {kluai}
\bibfield{author}{\bibinfo{person}{klu.ai}.} \bibinfo{year}{[n.\,d.]}\natexlab{}.
\newblock \bibinfo{title}{klu.ai}.
\newblock
\urldef\tempurl%
\url{https://docs.klu.ai/}
\showURL{%
\tempurl}


\bibitem[Li et~al\mbox{.}(2024a)]%
        {SSET}
\bibfield{author}{\bibinfo{person}{Muzhi Li}, \bibinfo{person}{Minda Hu}, \bibinfo{person}{Irwin King}, {and} \bibinfo{person}{Ho-fung Leung}.} \bibinfo{year}{2024}\natexlab{a}.
\newblock \showarticletitle{The Integration of Semantic and Structural Knowledge in Knowledge Graph Entity Typing}. In \bibinfo{booktitle}{\emph{Proceedings of the 2024 Conference of the North American Chapter of the Association for Computational Linguistics: Human Language Technologies (Volume 1: Long Papers)}}, \bibfield{editor}{\bibinfo{person}{Kevin Duh}, \bibinfo{person}{Helena Gomez}, {and} \bibinfo{person}{Steven Bethard}} (Eds.). \bibinfo{publisher}{Association for Computational Linguistics}, \bibinfo{address}{Mexico City, Mexico}, \bibinfo{pages}{6625--6638}.
\newblock
\href{https://doi.org/10.18653/v1/2024.naacl-long.369}{doi:\nolinkurl{10.18653/v1/2024.naacl-long.369}}


\bibitem[Li et~al\mbox{.}(2024b)]%
        {li2024matchinggenerationsurveygenerative}
\bibfield{author}{\bibinfo{person}{Xiaoxi Li}, \bibinfo{person}{Jiajie Jin}, \bibinfo{person}{Yujia Zhou}, \bibinfo{person}{Yuyao Zhang}, \bibinfo{person}{Peitian Zhang}, \bibinfo{person}{Yutao Zhu}, {and} \bibinfo{person}{Zhicheng Dou}.} \bibinfo{year}{2024}\natexlab{b}.
\newblock \bibinfo{title}{From Matching to Generation: A Survey on Generative Information Retrieval}.
\newblock
\showeprint[arxiv]{2404.14851}~[cs.IR]
\urldef\tempurl%
\url{https://arxiv.org/abs/2404.14851}
\showURL{%
\tempurl}


\bibitem[Liu(2022)]%
        {Liu_LlamaIndex_2022}
\bibfield{author}{\bibinfo{person}{Jerry Liu}.} \bibinfo{year}{2022}\natexlab{}.
\newblock \bibinfo{booktitle}{\emph{{LlamaIndex}}}.
\newblock
\href{https://doi.org/10.5281/zenodo.1234}{doi:\nolinkurl{10.5281/zenodo.1234}}


\bibitem[Liu et~al\mbox{.}(2023)]%
        {liu2023lostmiddlelanguagemodels}
\bibfield{author}{\bibinfo{person}{Nelson~F. Liu}, \bibinfo{person}{Kevin Lin}, \bibinfo{person}{John Hewitt}, \bibinfo{person}{Ashwin Paranjape}, \bibinfo{person}{Michele Bevilacqua}, \bibinfo{person}{Fabio Petroni}, {and} \bibinfo{person}{Percy Liang}.} \bibinfo{year}{2023}\natexlab{}.
\newblock \bibinfo{title}{Lost in the Middle: How Language Models Use Long Contexts}.
\newblock
\showeprint[arxiv]{2307.03172}~[cs.CL]
\urldef\tempurl%
\url{https://arxiv.org/abs/2307.03172}
\showURL{%
\tempurl}


\bibitem[Liu et~al\mbox{.}(2024)]%
        {liu2024raisf}
\bibfield{author}{\bibinfo{person}{Yanming Liu}, \bibinfo{person}{Xinyue Peng}, \bibinfo{person}{Xuhong Zhang}, \bibinfo{person}{Weihao Liu}, \bibinfo{person}{Jianwei Yin}, \bibinfo{person}{Jiannan Cao}, {and} \bibinfo{person}{Tianyu Du}.} \bibinfo{year}{2024}\natexlab{}.
\newblock \showarticletitle{RA-ISF: Learning to answer and understand from retrieval augmentation via iterative self-feedback}.
\newblock \bibinfo{journal}{\emph{arXiv preprint arXiv:2403.06840}} (\bibinfo{year}{2024}).
\newblock


\bibitem[Long et~al\mbox{.}(2024)]%
        {10.1609/aaai.v38i17.29837}
\bibfield{author}{\bibinfo{person}{Xinwei Long}, \bibinfo{person}{Jiali Zeng}, \bibinfo{person}{Fandong Meng}, \bibinfo{person}{Zhiyuan Ma}, \bibinfo{person}{Kaiyan Zhang}, \bibinfo{person}{Bowen Zhou}, {and} \bibinfo{person}{Jie Zhou}.} \bibinfo{year}{2024}\natexlab{}.
\newblock \showarticletitle{Generative multi-modal knowledge retrieval with large language models}. In \bibinfo{booktitle}{\emph{Proceedings of the Thirty-Eighth AAAI Conference on Artificial Intelligence and Thirty-Sixth Conference on Innovative Applications of Artificial Intelligence and Fourteenth Symposium on Educational Advances in Artificial Intelligence}} \emph{(\bibinfo{series}{AAAI'24/IAAI'24/EAAI'24})}. \bibinfo{publisher}{AAAI Press}, Article \bibinfo{articleno}{2089}, \bibinfo{numpages}{9}~pages.
\newblock
\showISBNx{978-1-57735-887-9}
\href{https://doi.org/10.1609/aaai.v38i17.29837}{doi:\nolinkurl{10.1609/aaai.v38i17.29837}}


\bibitem[Ma et~al\mbox{.}(2025)]%
        {ma2025thinkongraph}
\bibfield{author}{\bibinfo{person}{Shengjie Ma}, \bibinfo{person}{Chengjin Xu}, \bibinfo{person}{Xuhui Jiang}, \bibinfo{person}{Muzhi Li}, \bibinfo{person}{Huaren Qu}, \bibinfo{person}{Cehao Yang}, \bibinfo{person}{Jiaxin Mao}, {and} \bibinfo{person}{Jian Guo}.} \bibinfo{year}{2025}\natexlab{}.
\newblock \showarticletitle{Think-on-Graph 2.0: Deep and Faithful Large Language Model Reasoning with Knowledge-guided Retrieval Augmented Generation}. In \bibinfo{booktitle}{\emph{The Thirteenth International Conference on Learning Representations}}.
\newblock
\urldef\tempurl%
\url{https://openreview.net/forum?id=oFBu7qaZpS}
\showURL{%
\tempurl}


\bibitem[Moore et~al\mbox{.}(2023)]%
        {DBLP:journals/corr/abs-2307-08161}
\bibfield{author}{\bibinfo{person}{Steven Moore}, \bibinfo{person}{Huy~Anh Nguyen}, \bibinfo{person}{Tianying Chen}, {and} \bibinfo{person}{John~C. Stamper}.} \bibinfo{year}{2023}\natexlab{}.
\newblock \showarticletitle{Assessing the Quality of Multiple-Choice Questions Using {GPT-4} and Rule-Based Methods}.
\newblock \bibinfo{journal}{\emph{CoRR}}  \bibinfo{volume}{abs/2307.08161} (\bibinfo{year}{2023}).
\newblock
\href{https://doi.org/10.48550/ARXIV.2307.08161}{doi:\nolinkurl{10.48550/ARXIV.2307.08161}}
\showeprint[arXiv]{2307.08161}


\bibitem[Naismith et~al\mbox{.}(2023)]%
        {naismith-etal-2023-automated}
\bibfield{author}{\bibinfo{person}{Ben Naismith}, \bibinfo{person}{Phoebe Mulcaire}, {and} \bibinfo{person}{Jill Burstein}.} \bibinfo{year}{2023}\natexlab{}.
\newblock \showarticletitle{Automated evaluation of written discourse coherence using {GPT}-4}. In \bibinfo{booktitle}{\emph{Proceedings of the 18th Workshop on Innovative Use of NLP for Building Educational Applications (BEA 2023)}}, \bibfield{editor}{\bibinfo{person}{Ekaterina Kochmar}, \bibinfo{person}{Jill Burstein}, \bibinfo{person}{Andrea Horbach}, \bibinfo{person}{Ronja Laarmann-Quante}, \bibinfo{person}{Nitin Madnani}, \bibinfo{person}{Ana{\"i}s Tack}, \bibinfo{person}{Victoria Yaneva}, \bibinfo{person}{Zheng Yuan}, {and} \bibinfo{person}{Torsten Zesch}} (Eds.). \bibinfo{publisher}{Association for Computational Linguistics}, \bibinfo{address}{Toronto, Canada}, \bibinfo{pages}{394--403}.
\newblock
\href{https://doi.org/10.18653/v1/2023.bea-1.32}{doi:\nolinkurl{10.18653/v1/2023.bea-1.32}}


\bibitem[Rajput et~al\mbox{.}(2023)]%
        {rajput2023recommendersystemsgenerativeretrieval}
\bibfield{author}{\bibinfo{person}{Shashank Rajput}, \bibinfo{person}{Nikhil Mehta}, \bibinfo{person}{Anima Singh}, \bibinfo{person}{Raghunandan~H. Keshavan}, \bibinfo{person}{Trung Vu}, \bibinfo{person}{Lukasz Heldt}, \bibinfo{person}{Lichan Hong}, \bibinfo{person}{Yi Tay}, \bibinfo{person}{Vinh~Q. Tran}, \bibinfo{person}{Jonah Samost}, \bibinfo{person}{Maciej Kula}, \bibinfo{person}{Ed~H. Chi}, {and} \bibinfo{person}{Maheswaran Sathiamoorthy}.} \bibinfo{year}{2023}\natexlab{}.
\newblock \bibinfo{title}{Recommender Systems with Generative Retrieval}.
\newblock
\showeprint[arxiv]{2305.05065}~[cs.IR]
\urldef\tempurl%
\url{https://arxiv.org/abs/2305.05065}
\showURL{%
\tempurl}


\bibitem[Riedler and Langer(2024)]%
        {riedler2024textoptimizingragmultimodal}
\bibfield{author}{\bibinfo{person}{Monica Riedler} {and} \bibinfo{person}{Stefan Langer}.} \bibinfo{year}{2024}\natexlab{}.
\newblock \bibinfo{title}{Beyond Text: Optimizing RAG with Multimodal Inputs for Industrial Applications}.
\newblock
\showeprint[arxiv]{2410.21943}~[cs.CL]
\urldef\tempurl%
\url{https://arxiv.org/abs/2410.21943}
\showURL{%
\tempurl}


\bibitem[Robertson et~al\mbox{.}(2009)]%
        {robertson2009probabilistic}
\bibfield{author}{\bibinfo{person}{Stephen Robertson}, \bibinfo{person}{Hugo Zaragoza}, {et~al\mbox{.}}} \bibinfo{year}{2009}\natexlab{}.
\newblock \showarticletitle{The probabilistic relevance framework: BM25 and beyond}.
\newblock \bibinfo{journal}{\emph{Foundations and Trends{\textregistered} in Information Retrieval}} \bibinfo{volume}{3}, \bibinfo{number}{4} (\bibinfo{year}{2009}), \bibinfo{pages}{333--389}.
\newblock


\bibitem[Schwartz(2021)]%
        {schwartz2021ensemblemrrndcgmodels}
\bibfield{author}{\bibinfo{person}{Idan Schwartz}.} \bibinfo{year}{2021}\natexlab{}.
\newblock \bibinfo{title}{Ensemble of MRR and NDCG models for Visual Dialog}.
\newblock
\showeprint[arxiv]{2104.07511}~[cs.AI]
\urldef\tempurl%
\url{https://arxiv.org/abs/2104.07511}
\showURL{%
\tempurl}


\bibitem[SEC(2024a)]%
        {jpm-sec}
\bibfield{author}{\bibinfo{person}{SEC}.} \bibinfo{year}{2024}\natexlab{a}.
\newblock \bibinfo{title}{JP Morgan Affiliates to Pay 151 Million to Resolve SEC Enforcement Actions}.
\newblock
\urldef\tempurl%
\url{https://www.sec.gov/newsroom/press-releases/2024-178}
\showURL{%
\tempurl}


\bibitem[SEC(2024b)]%
        {gs-sec}
\bibfield{author}{\bibinfo{person}{SEC}.} \bibinfo{year}{2024}\natexlab{b}.
\newblock \bibinfo{title}{SEC Levies More Than 3.8 Million in Penalties in Sweep of Late Beneficial Ownership and Insider Transaction Reports}.
\newblock
\urldef\tempurl%
\url{https://www.sec.gov/newsroom/press-releases/2024-148}
\showURL{%
\tempurl}


\bibitem[Setty et~al\mbox{.}(2024)]%
        {setty2024improvingretrievalragbased}
\bibfield{author}{\bibinfo{person}{Spurthi Setty}, \bibinfo{person}{Harsh Thakkar}, \bibinfo{person}{Alyssa Lee}, \bibinfo{person}{Eden Chung}, {and} \bibinfo{person}{Natan Vidra}.} \bibinfo{year}{2024}\natexlab{}.
\newblock \bibinfo{title}{Improving Retrieval for RAG based Question Answering Models on Financial Documents}.
\newblock
\showeprint[arxiv]{2404.07221}~[cs.IR]
\urldef\tempurl%
\url{https://arxiv.org/abs/2404.07221}
\showURL{%
\tempurl}


\bibitem[Shah et~al\mbox{.}(2024)]%
        {shah2024multidocumentfinancialquestionanswering}
\bibfield{author}{\bibinfo{person}{Shalin Shah}, \bibinfo{person}{Srikanth Ryali}, {and} \bibinfo{person}{Ramasubbu Venkatesh}.} \bibinfo{year}{2024}\natexlab{}.
\newblock \bibinfo{title}{Multi-Document Financial Question Answering using LLMs}.
\newblock
\showeprint[arxiv]{2411.07264}~[cs.IR]
\urldef\tempurl%
\url{https://arxiv.org/abs/2411.07264}
\showURL{%
\tempurl}


\bibitem[Sultania et~al\mbox{.}(2024)]%
        {sultania2024domainspecific}
\bibfield{author}{\bibinfo{person}{Dewang Sultania}, \bibinfo{person}{Zhaoyu Lu}, \bibinfo{person}{Twisha Naik}, \bibinfo{person}{Franck Dernoncourt}, \bibinfo{person}{David~Seunghyun Yoon}, \bibinfo{person}{Sanat Sharma}, \bibinfo{person}{Trung Bui}, \bibinfo{person}{Ashok Gupta}, \bibinfo{person}{Tushar Vatsa}, \bibinfo{person}{Suhas Suresha}, \bibinfo{person}{Ishita Verma}, \bibinfo{person}{Vibha Belavadi}, \bibinfo{person}{Cheng Chen}, {and} \bibinfo{person}{Michael Friedrich}.} \bibinfo{year}{2024}\natexlab{}.
\newblock \bibinfo{title}{Domain-specific Question Answering with Hybrid Search}.
\newblock
\showeprint[arxiv]{2412.03736}~[cs.CL]
\urldef\tempurl%
\url{https://arxiv.org/abs/2412.03736}
\showURL{%
\tempurl}


\bibitem[Wang et~al\mbox{.}(2024)]%
        {wang2024biorag}
\bibfield{author}{\bibinfo{person}{C. Wang}, \bibinfo{person}{Q. Long}, \bibinfo{person}{M. Xiao}, {et~al\mbox{.}}} \bibinfo{year}{2024}\natexlab{}.
\newblock \showarticletitle{Biorag: A RAG-LLM framework for biological question reasoning}.
\newblock \bibinfo{journal}{\emph{arXiv preprint arXiv:2408.01107}} (\bibinfo{year}{2024}).
\newblock


\bibitem[Wang et~al\mbox{.}(2023)]%
        {wang2023selfknowledge}
\bibfield{author}{\bibinfo{person}{Yile Wang}, \bibinfo{person}{Peng Li}, \bibinfo{person}{Maosong Sun}, {and} \bibinfo{person}{Yang Liu}.} \bibinfo{year}{2023}\natexlab{}.
\newblock \showarticletitle{Self-knowledge guided retrieval augmentation for large language models}. In \bibinfo{booktitle}{\emph{Findings of the Association for Computational Linguistics: EMNLP 2023}}. \bibinfo{pages}{10303--10315}.
\newblock


\bibitem[Xia et~al\mbox{.}(2024)]%
        {xia2024mmedragversatilemultimodalrag}
\bibfield{author}{\bibinfo{person}{Peng Xia}, \bibinfo{person}{Kangyu Zhu}, \bibinfo{person}{Haoran Li}, \bibinfo{person}{Tianze Wang}, \bibinfo{person}{Weijia Shi}, \bibinfo{person}{Sheng Wang}, \bibinfo{person}{Linjun Zhang}, \bibinfo{person}{James Zou}, {and} \bibinfo{person}{Huaxiu Yao}.} \bibinfo{year}{2024}\natexlab{}.
\newblock \bibinfo{title}{MMed-RAG: Versatile Multimodal RAG System for Medical Vision Language Models}.
\newblock
\showeprint[arxiv]{2410.13085}~[cs.LG]
\urldef\tempurl%
\url{https://arxiv.org/abs/2410.13085}
\showURL{%
\tempurl}


\bibitem[Xie et~al\mbox{.}(2024)]%
        {xie2024openfinllmsopenmultimodallarge}
\bibfield{author}{\bibinfo{person}{Qianqian Xie}, \bibinfo{person}{Dong Li}, \bibinfo{person}{Mengxi Xiao}, \bibinfo{person}{Zihao Jiang}, \bibinfo{person}{Ruoyu Xiang}, \bibinfo{person}{Xiao Zhang}, \bibinfo{person}{Zhengyu Chen}, \bibinfo{person}{Yueru He}, \bibinfo{person}{Weiguang Han}, \bibinfo{person}{Yuzhe Yang}, \bibinfo{person}{Shunian Chen}, \bibinfo{person}{Yifei Zhang}, \bibinfo{person}{Lihang Shen}, \bibinfo{person}{Daniel Kim}, \bibinfo{person}{Zhiwei Liu}, \bibinfo{person}{Zheheng Luo}, \bibinfo{person}{Yangyang Yu}, \bibinfo{person}{Yupeng Cao}, \bibinfo{person}{Zhiyang Deng}, \bibinfo{person}{Zhiyuan Yao}, \bibinfo{person}{Haohang Li}, \bibinfo{person}{Duanyu Feng}, \bibinfo{person}{Yongfu Dai}, \bibinfo{person}{VijayaSai Somasundaram}, \bibinfo{person}{Peng Lu}, \bibinfo{person}{Yilun Zhao}, \bibinfo{person}{Yitao Long}, \bibinfo{person}{Guojun Xiong}, \bibinfo{person}{Kaleb Smith}, \bibinfo{person}{Honghai Yu}, \bibinfo{person}{Yanzhao Lai}, \bibinfo{person}{Min Peng},
  \bibinfo{person}{Jianyun Nie}, \bibinfo{person}{Jordan~W. Suchow}, \bibinfo{person}{Xiao-Yang Liu}, \bibinfo{person}{Benyou Wang}, \bibinfo{person}{Alejandro Lopez-Lira}, \bibinfo{person}{Jimin Huang}, {and} \bibinfo{person}{Sophia Ananiadou}.} \bibinfo{year}{2024}\natexlab{}.
\newblock \bibinfo{title}{Open-FinLLMs: Open Multimodal Large Language Models for Financial Applications}.
\newblock
\showeprint[arxiv]{2408.11878}~[cs.CL]
\urldef\tempurl%
\url{https://arxiv.org/abs/2408.11878}
\showURL{%
\tempurl}


\bibitem[Xiong et~al\mbox{.}(2020)]%
        {xiong2020approximate}
\bibfield{author}{\bibinfo{person}{Lee Xiong}, \bibinfo{person}{Chenyan Xiong}, \bibinfo{person}{Ye Li}, \bibinfo{person}{Kwok-Fung Tang}, \bibinfo{person}{Jialin Liu}, \bibinfo{person}{Paul Bennett}, \bibinfo{person}{Junaid Ahmed}, {and} \bibinfo{person}{Arnold Overwijk}.} \bibinfo{year}{2020}\natexlab{}.
\newblock \showarticletitle{Approximate nearest neighbor negative contrastive learning for dense text retrieval}.
\newblock \bibinfo{journal}{\emph{arXiv preprint arXiv:2007.00808}} (\bibinfo{year}{2020}).
\newblock


\bibitem[Yu et~al\mbox{.}(2025)]%
        {yu2025visrag}
\bibfield{author}{\bibinfo{person}{Shi Yu}, \bibinfo{person}{Chaoyue Tang}, \bibinfo{person}{Bokai Xu}, \bibinfo{person}{Junbo Cui}, \bibinfo{person}{Junhao Ran}, \bibinfo{person}{Yukun Yan}, \bibinfo{person}{Zhenghao Liu}, \bibinfo{person}{Shuo Wang}, \bibinfo{person}{Xu Han}, \bibinfo{person}{Zhiyuan Liu}, {and} \bibinfo{person}{Maosong Sun}.} \bibinfo{year}{2025}\natexlab{}.
\newblock \showarticletitle{Vis{RAG}: Vision-based Retrieval-augmented Generation on Multi-modality Documents}. In \bibinfo{booktitle}{\emph{The Thirteenth International Conference on Learning Representations}}.
\newblock
\urldef\tempurl%
\url{https://openreview.net/forum?id=zG459X3Xge}
\showURL{%
\tempurl}


\bibitem[Yuan et~al\mbox{.}(2024)]%
        {yuan2024hybridragcomprehensiveenhancement}
\bibfield{author}{\bibinfo{person}{Ye Yuan}, \bibinfo{person}{Chengwu Liu}, \bibinfo{person}{Jingyang Yuan}, \bibinfo{person}{Gongbo Sun}, \bibinfo{person}{Siqi Li}, {and} \bibinfo{person}{Ming Zhang}.} \bibinfo{year}{2024}\natexlab{}.
\newblock \bibinfo{title}{A Hybrid RAG System with Comprehensive Enhancement on Complex Reasoning}.
\newblock
\showeprint[arxiv]{2408.05141}~[cs.CL]
\urldef\tempurl%
\url{https://arxiv.org/abs/2408.05141}
\showURL{%
\tempurl}


\bibitem[Zhai(2024)]%
        {zhai2024selfadaptivemultimodalretrievalaugmentedgeneration}
\bibfield{author}{\bibinfo{person}{Wenjia Zhai}.} \bibinfo{year}{2024}\natexlab{}.
\newblock \bibinfo{title}{Self-adaptive Multimodal Retrieval-Augmented Generation}.
\newblock
\showeprint[arxiv]{2410.11321}~[cs.CL]
\urldef\tempurl%
\url{https://arxiv.org/abs/2410.11321}
\showURL{%
\tempurl}


\bibitem[Zhang et~al\mbox{.}(2023)]%
        {zhang2023moqagpt}
\bibfield{author}{\bibinfo{person}{Le Zhang}, \bibinfo{person}{Yihong Wu}, \bibinfo{person}{Fengran Mo}, \bibinfo{person}{Jian-Yun Nie}, {and} \bibinfo{person}{Aishwarya Agrawal}.} \bibinfo{year}{2023}\natexlab{}.
\newblock \showarticletitle{MoqaGPT: Zero-Shot Multi-modal Open-domain Question Answering with Large Language Model}.
\newblock \bibinfo{journal}{\emph{arXiv preprint arXiv:2310.13265}} (\bibinfo{year}{2023}).
\newblock


\bibitem[Zhang et~al\mbox{.}(2024b)]%
        {zhang2024exploring}
\bibfield{author}{\bibinfo{person}{Le Zhang}, \bibinfo{person}{Yihong Wu}, \bibinfo{person}{Qian Yang}, {and} \bibinfo{person}{Jian-Yun Nie}.} \bibinfo{year}{2024}\natexlab{b}.
\newblock \showarticletitle{Exploring the Best Practices of Query Expansion with Large Language Models}. In \bibinfo{booktitle}{\emph{Findings of the Association for Computational Linguistics: EMNLP 2024}}. \bibinfo{pages}{1872--1883}.
\newblock


\bibitem[Zhang et~al\mbox{.}(2024a)]%
        {zhang2024instructiontuninglargelanguage}
\bibfield{author}{\bibinfo{person}{Shengyu Zhang}, \bibinfo{person}{Linfeng Dong}, \bibinfo{person}{Xiaoya Li}, \bibinfo{person}{Sen Zhang}, \bibinfo{person}{Xiaofei Sun}, \bibinfo{person}{Shuhe Wang}, \bibinfo{person}{Jiwei Li}, \bibinfo{person}{Runyi Hu}, \bibinfo{person}{Tianwei Zhang}, \bibinfo{person}{Fei Wu}, {and} \bibinfo{person}{Guoyin Wang}.} \bibinfo{year}{2024}\natexlab{a}.
\newblock \bibinfo{title}{Instruction Tuning for Large Language Models: A Survey}.
\newblock
\showeprint[arxiv]{2308.10792}~[cs.CL]
\urldef\tempurl%
\url{https://arxiv.org/abs/2308.10792}
\showURL{%
\tempurl}


\bibitem[Zhao et~al\mbox{.}(2024)]%
        {zhao2024longragdualperspectiveretrievalaugmentedgeneration}
\bibfield{author}{\bibinfo{person}{Qingfei Zhao}, \bibinfo{person}{Ruobing Wang}, \bibinfo{person}{Yukuo Cen}, \bibinfo{person}{Daren Zha}, \bibinfo{person}{Shicheng Tan}, \bibinfo{person}{Yuxiao Dong}, {and} \bibinfo{person}{Jie Tang}.} \bibinfo{year}{2024}\natexlab{}.
\newblock \bibinfo{title}{LongRAG: A Dual-Perspective Retrieval-Augmented Generation Paradigm for Long-Context Question Answering}.
\newblock
\showeprint[arxiv]{2410.18050}~[cs.CL]
\urldef\tempurl%
\url{https://arxiv.org/abs/2410.18050}
\showURL{%
\tempurl}


\bibitem[Zheng et~al\mbox{.}(2023)]%
        {zheng2023judgingllmasajudgemtbenchchatbot}
\bibfield{author}{\bibinfo{person}{Lianmin Zheng}, \bibinfo{person}{Wei-Lin Chiang}, \bibinfo{person}{Ying Sheng}, \bibinfo{person}{Siyuan Zhuang}, \bibinfo{person}{Zhanghao Wu}, \bibinfo{person}{Yonghao Zhuang}, \bibinfo{person}{Zi Lin}, \bibinfo{person}{Zhuohan Li}, \bibinfo{person}{Dacheng Li}, \bibinfo{person}{Eric~P. Xing}, \bibinfo{person}{Hao Zhang}, \bibinfo{person}{Joseph~E. Gonzalez}, {and} \bibinfo{person}{Ion Stoica}.} \bibinfo{year}{2023}\natexlab{}.
\newblock \bibinfo{title}{Judging LLM-as-a-Judge with MT-Bench and Chatbot Arena}.
\newblock
\showeprint[arxiv]{2306.05685}~[cs.CL]
\urldef\tempurl%
\url{https://arxiv.org/abs/2306.05685}
\showURL{%
\tempurl}


\bibitem[Zhu et~al\mbox{.}(2024)]%
        {zhu2024rageval}
\bibfield{author}{\bibinfo{person}{Kunlun Zhu}, \bibinfo{person}{Yifan Luo}, \bibinfo{person}{Dingling Xu}, \bibinfo{person}{Ruobing Wang}, \bibinfo{person}{Shi Yu}, \bibinfo{person}{Shuo Wang}, \bibinfo{person}{Yukun Yan}, \bibinfo{person}{Zhenghao Liu}, \bibinfo{person}{Xu Han}, \bibinfo{person}{Zhiyuan Liu}, {et~al\mbox{.}}} \bibinfo{year}{2024}\natexlab{}.
\newblock \showarticletitle{Rageval: Scenario specific rag evaluation dataset generation framework}.
\newblock \bibinfo{journal}{\emph{arXiv preprint arXiv:2408.01262}} (\bibinfo{year}{2024}).
\newblock


\end{thebibliography}

% %%
% %% If your work has an appendix, this is the place to put it.
% \appendix






\newpage
% \onecolumn % Switch to single-column mode
\appendix
\begin{table*}[h!]
    \centering
    \caption{Retrieval Performance on the Company Dataset with a Fixed top-$K$. Avg Num Retrieved represents the average number of chunks in the retrieved set. Top‑K is set to 10, so each retriever selects its top 10 results for the retrieval set.}
    \label{tab:}
    \resizebox{\textwidth}{!}{ 
    \begin{tabular}{p{3.8cm} p{5.5cm} cccc}
        \toprule
        \small \textbf{Category} & \textbf{Method} & \textbf{Avg Recall} & \textbf{Avg Precision} & \textbf{Avg F1} & \textbf{Avg Num Retrieved} \\
        \midrule
        \multirow{2}{*}{\small \textbf{FAISS Retrieval}} 
        & FAISS (Baseline) & 0.5984 & 0.4427 & 0.4861 & 10.8 \\ % Standard FAISS retrieval
        & \quad + Bundle Expansion(Exp) & 0.6206 & \textbf{0.4440} & \textbf{0.4960} & 11.28 \\ % Adds chunk expansion
        \midrule
        \multirow{3}{*}{\small \textbf{HyDE Retrieval}} 
        & \quad \textbf{HyDE-1}: HyDE(Qwen7B, w/ Exp) & 0.7019 & 0.2119 & 0.3084 & 28.73 \\ % Base HyDE
        & \quad \textbf{HyDE-2}: HyDE(Qwen7B-SFT, w/ Exp) & 0.7186 & 0.2438 & 0.3450 & 25.48 \\ % Fine-tuned HyDE (Used in our model)
        & \quad \textbf{HyDE-3}: HyDE(Qwen72B, w/ Exp) & 0.7310 & 0.2160 & 0.3178 & 29.49 \\ % Large model, but no fine-tuning
        \midrule
        \small \textbf{\ourmodel{}} & \quad + BM25+Metadata+\textbf{HyDE-2} & \textbf{0.9251} & 0.1272 & 0.2156 & 68.75 \\ % Our best configuration
        \bottomrule
    \end{tabular}
    }
\end{table*}

\section{Data Preprocessing Details}
To effectively convert PDF files into structured data, we utilize the Mineru tool~\footnote{https://github.com/opendatalab/MinerU}, which can extract content from PDFs and export it in various formats. The preprocessing workflow is described below.

\subsection{PDF to JSON Conversion Process}

We can utilize the Mineru to to process PDF files with the following command:

\begin{verbatim}
magic-pdf -p {some_pdf} -o {some_output_dir} -m auto
\end{verbatim}

Users may need to specify the input PDF file and the output directory. The tool generates the following output files:

\begin{itemize} [leftmargin=*]
    \item \texttt{some\_pdf.md}: Markdown representation of the PDF content.
    \item \texttt{images/}: A directory containing extracted images, including tables.
    \item \texttt{some\_pdf\_layout.pdf}: A visualization of the document's layout.
    \item \texttt{some\_pdf\_middle.json}: Intermediate processing results in JSON format.
    \item \texttt{some\_pdf\_model.json}: Model inference results in JSON format.
    \item \texttt{some\_pdf\_origin.pdf}: The original PDF file.
    \item \texttt{some\_pdf\_spans.pdf}: A PDF file showing the bounding box positions at the smallest granularity.
    \item \texttt{some\_pdf\_content\_list.json}: A structured JSON file representing the document content in reading order.
\end{itemize}

Among these files, \( \texttt{some\_pdf\_content\_list.json} \) stores the document content into blocks, each of which contains the following fields: \( \texttt{type} \), \( \texttt{text} \), \( \texttt{text\_level} \), and \( \texttt{page\_idx} \). The \( \texttt{type} \) field differentiates different content types, such as:

- \( \texttt{text} \): Regular text blocks.
- \( \texttt{table} \): Table elements.
- \( \texttt{img\_path} \): Image paths referencing extracted tables.

\subsection{Examples of JSON Structure}

\textbf{Text Block Example:}
\begin{figure}[H]
    \centering
    \includegraphics[width=1\linewidth]{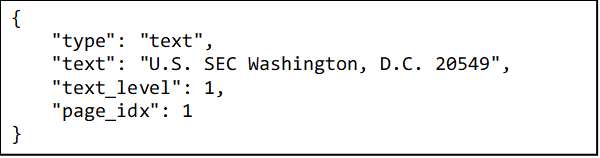}
    \caption*{In this example, \( \texttt{type} = \texttt{text} \) signifies a text block, while \( \texttt{text\_level} = 1 \) designates it as a title.}
\end{figure}
%\begin{verbatim}
%{
%    "type": "text",
%    "text": "U.S. SEC Washington, D.C. 20549",
%    "text_level": 1,
%    "page_idx": 1
%}
%\end{verbatim}

% \begin{figure}
%     \centering
%     \includegraphics[width=0.5\linewidth]{text_block_example.png}
%     \caption{Enter Caption}
%     \label{fig:enter-label}
%     \Description{Table Block Example}
% \end{figure}
%In this example, \( \texttt{type} = \texttt{text} \) signifies a text block, while \( \texttt{text\_level} = 1 \) designates it as a title.

\textbf{Table Block Example:}
\begin{figure}[H]
    \centering
    \includegraphics[width=1\linewidth]{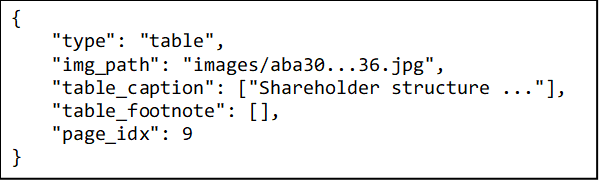}
    \caption*{Here, \( \texttt{type} = \texttt{table} \) denotes a table, with \( \texttt{img\_path} \) pointing to the extracted table image.}
\end{figure}
%\begin{verbatim}
%{
%    "type": "table",
%    "img_path": "images/aba30...36.jpg",
%    "table_caption": ["Shareholder structure ..."],
%    "table_footnote": [],
%    "page_idx": 9
%}
%\end{verbatim}
%Here, \( \texttt{type} = \texttt{table} \) denotes a table, with \( \texttt{img\_path} \) pointing to the extracted table image.

\textbf{Image Block Example:}
\begin{figure}[H]
    \centering
    \includegraphics[width=1\linewidth]{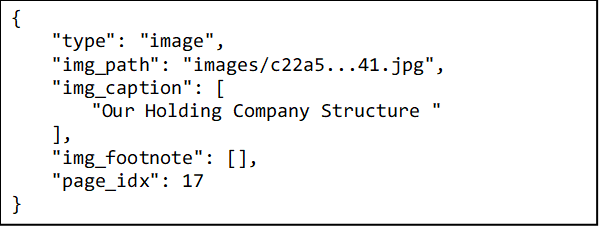}
    \caption*{Here, \( \texttt{type} = \texttt{image} \) denotes a image, with \( \texttt{img\_path} \) pointing to the extracted image.}
\end{figure}
%\begin{verbatim}
%{
%    "type": "image",
%    "img_path": "images/c22a5...41.jpg",
%    "img_caption": [
%        "Our Holding Company Structure "
%    ],
%    "img_footnote": [],
%    "page_idx": 17
%}
%\end{verbatim}
%Here, \( \texttt{type} = \texttt{image} \) denotes a image, with \( \texttt{img\_path} \) pointing to the extracted image.

\subsection{Image Processing and Text Parsing}
\label{appx:vlm}
To extract table information, the Mineru2Base tool leverages the GPT-4o API to convert table images into text. The system iterates through each document block in  \texttt{some\_pdf\_content\_list.json} . For blocks containing an  \texttt{img\_path} , the tool retrieves the preceding and succeeding blocks as context. The table image is then converted to Base64 format and passed to the GPT-4o API, which generates a structured textual representation.
\subsection{Embedding Model}
\label{appx:embedding_model}

Two embedding models are used in this project, each serving a distinct purpose: TF-IDF for text deduplication and BGE-M3 for storing text in the database.

For the deduplication task, we employ the TF-IDF tool to convert the text into vector representations.  This approach helps to identify and remove redundant text segments by comparing their vector similarities.  The model uses term frequency and inverse document frequency to generate vectors, which are then compared using cosine similarity to detect high similarity between text segments.  Text with similarities above a predefined threshold is removed, reducing redundancy.

For storing text in the Chroma database, we use the BGE-M3 embedding model.  This model, trained on large-scale datasets, transforms text into high-dimensional vectors that capture the semantic meaning of the content.  These vectors are then stored in the database, enabling efficient indexing and retrieval for future similarity searches and semantic analysis.
\subsection{Duplicate Removal}

To remove redundant content, a de-duplication step is performed using text similarity measurements. The method includes:

1.Chinese word segmentation using the Jieba library.
2.ransforming text into a term-frequency matrix via PyTorch, where the text is vectorized using a TF-IDF approach implemented with torch for efficient matrix computation.
3.Computing cosine similarity between document blocks using PyTorch’s cosine similarity function.

If the similarity score exceeds 0.7 (default threshold), the duplicate block is removed.

\subsection{Coreference Resolution}

Coreference resolution is an important task in natural language processing, aiming to identify the entities that pronouns and other referring expressions point to within a text. In simple terms, it analyzes pronouns (such as ``he," ``it," ``we," etc.) and determines the specific object or entity they refer to, making the text more coherent and clear. The relationship between pronouns and their referred entities can significantly affect comprehension, and thus coreference resolution plays a crucial role in improving text readability and reducing ambiguity.

In practical implementation, the algorithm traverses a JSON file containing multiple chunks, each with a title and content field. For each chunk, the coreference resolution follows a specific process: it first searches up to four preceding chunks as references. The search is limited to preceding chunks that share the same title as the current one. If the previous chunk belongs to a different title, it is considered to have low relevance to the current chunk and is thus not considered as a reference.

In this way, the coreference resolution algorithm effectively scans the related content within the text, using the GPT-4o-mini API to resolve pronoun references and replace them with specific entities. This approach not only improves the clarity and coherence of the text but also reduces the ambiguity that pronouns may cause, ensuring the accurate transmission of information.

Detail prompts provided to GPT-4o-mini during the anaphora resolution step are shown in Figure~\ref{fig:anaphoraresolution}:

\begin{figure*}
    \centering
    \includegraphics[width=1\linewidth]{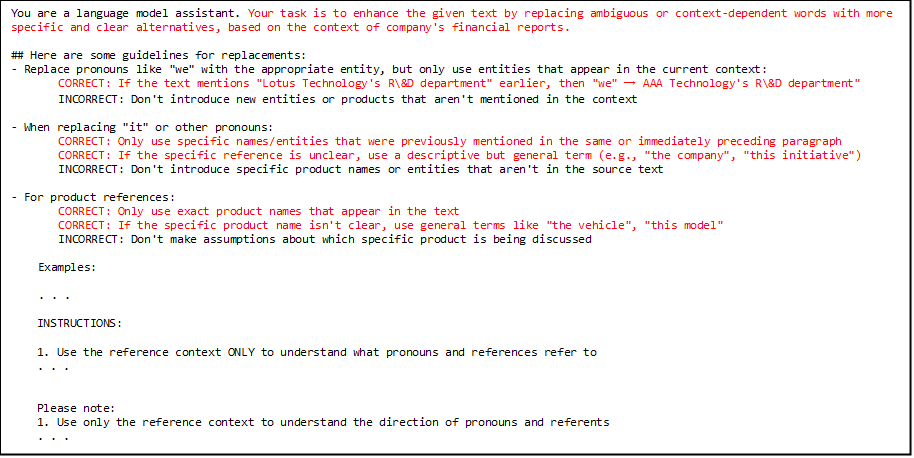}
    \caption{The instruction prompts for anaphora resolution.}
    \label{fig:anaphoraresolution}
\end{figure*}

\subsection{Title Summarization}

Each block receives a summary stored in the \( \texttt{title\_summary} \) field. The process involves:

1. Generating a concise summary for each block.
2. Aggregating summaries to form a higher-level title summary.
3. Storing the final summary in \( \texttt{title\_summary} \).

For example, the original JSON format:

\begin{figure}[H]
    \centering
    \includegraphics[width=1\linewidth]{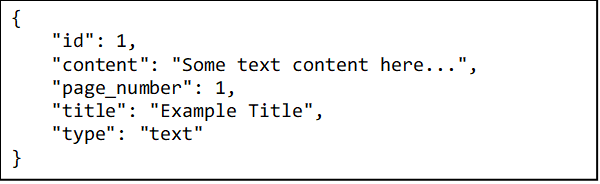}
\end{figure}
%\begin{verbatim}
%{
%    "id": 1,
%    "content": "Some text content here...",
%    "page_number": 1,
%    "title": "Example Title",
%    "type": "text"
%}
%\end{verbatim}

After summarization:
\begin{figure}[H]
    \centering
    \includegraphics[width=1\linewidth]{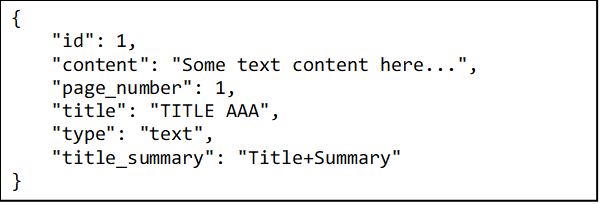}
\end{figure}
%\begin{verbatim}
%{
%    "id": 1,
%    "content": "Some text content here...",
%    "page_number": 1,
%    "title": "TITLE AAA",
%    "type": "text",
%    "title_summary": "Title+Summary"
%}
%\end{verbatim}

\subsection{Text Segmentation}

To maintain manageable block sizes, text segmentation is applied. The segmentation rules are:

- If a block has fewer than 200 characters, subsequent content is appended until it exceeds 200.
- If a block exceeds 200 characters, the last sentence is moved to the next block.
- Segmentation is restricted within the same title to prevent cross-title segmentation.

Through this structured preprocessing pipeline, PDF content is efficiently converted into JSON format, ensuring accurate extraction of text, tables, and other elements. This process facilitates subsequent analysis, model applications, and further information retrieval.

\subsection{Loading Chunk into a Vector Database}

In this study, text data is loaded into a Chroma-based vector database, using the BGE-M3 embedding model to vectorize the text.  The process begins by reading a series of JSON-format files from a specified directory, each containing metadata related to page content and the text itself.  Metadata such as page range and publication date are extracted and processed along with the corresponding content.  To avoid duplicate data, each piece of text is uniquely identified by calculating its SHA-256 hash, and in case of duplicate content, the system updates the record with the latest version based on the publication date.

Next, the text and associated metadata are stored into the Chroma database in batches.  During the storage process, the system also maintains the sequential relationship between documents by adding ``previous" and ``next" chunk identifiers for texts within the same file.  Finally, to further optimize information retrieval, a BM25 index is built based on the text data stored in the Chroma database, improving retrieval efficiency in subsequent searches.

Through this process, we can effectively convert large amounts of text data into vectors and provide efficient database support for future retrieval and analysis.

\section{Reranker Finetune Details}

\subsection{Data Preparation}
The training data is stored in a JSONL format, where each entry contains the following fields: \texttt{query}, \texttt{pos}, \texttt{neg}, and optionally, \texttt{pos\_scores}, \texttt{neg\_scores}, and \texttt{prompt}. The key fields are as follows:

\begin{itemize}[leftmargin=*]
    \item \texttt{query}: The query string.
    \item \texttt{pos}: A list containing one positive candidate related to the query.
    \item \texttt{neg}: A list containing multiple negative candidates unrelated to the query.
    \item \texttt{pos\_scores} (optional): A list of scores corresponding to the positive candidates.
    \item \texttt{neg\_scores} (optional): A list of scores corresponding to the negative candidates.
    \item \texttt{prompt} (optional): A prompt string used for additional customization of the training data.
\end{itemize}

The minimal required data format is as follows:

\begin{figure}[H]
    \centering
    \includegraphics[width=1\linewidth]{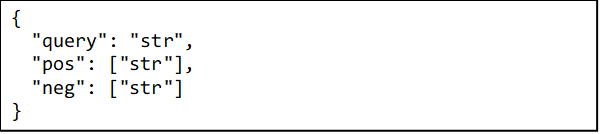}
\end{figure}
%\begin{lstlisting}[language=python]
%{
%  "query": "str", 
%  "pos": ["str"], 
%  "neg": ["str"]
%}
%\end{lstlisting}

In each labeled sample, the pos field is structured as a list containing only a single positive instance. This positive instance typically represents the most relevant or ideal answer in relation to the query, thereby providing the model with a clear and representative positive reference. In contrast, the neg field consists of a list of multiple negative candidate strings, each reflecting content that is either irrelevant or only weakly related to the query. This data construction facilitates a significant enhancement of the model’s sensitivity and discriminative capability in assessing relevance, achieved through contrastive learning between positive and negative samples during the fine-tuning of the reranker model.

Example of the labeled training data:
\begin{figure}[H]
    \centering
    \includegraphics[width=1\linewidth]{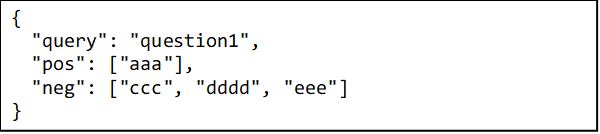}
\end{figure}
\begin{figure}[H]
    \centering
    \includegraphics[width=1\linewidth]{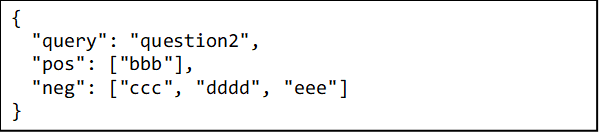}
\end{figure}
%\begin{lstlisting}[language=python]
%{
%  "query": "question1", 
%  "pos": ["aaa"], 
%  "neg": ["ccc", "dddd", "eee"]
%}
%{
%  "query": "question2", 
%  "pos": ["bbb"], 
%  "neg": ["ccc", "dddd", "eee"]
%}
%\end{lstlisting}

The above format is used for fine-tuning a reranker model, where the goal is to reorder the candidate passages based on their relevance to the query.

\subsection{Selection of Positive and Negative Samples}

In this study, the selection of positive and negative samples is governed by a systematic and rigorous procedure, meticulously designed to ensure that the chosen samples effectively facilitate both the training and subsequent optimization of the model. The process comprises several carefully orchestrated stages, which are detailed as follows:

Initially, a comprehensive set of questions was curated to represent a wide array of potential query scenarios that the model might encounter in real-world applications. These questions were selected to capture the inherent diversity and complexity of user inquiries, thereby ensuring that the dataset is both representative and robust. For each question, a dual retrieval strategy was implemented using the Faiss and BM25 algorithms within the Chroma vector database. Faiss, renowned for its efficiency in high-dimensional vector similarity search, and BM25, a well-established probabilistic retrieval model, complement each other by capturing both semantic nuances and lexical patterns. This combined approach ensures that each query retrieves a diverse set of candidate answers—typically exceeding twenty per question—thus providing a rich pool of potential responses with varying degrees of relevance and quality.

Subsequent to the retrieval phase, the candidate results underwent an exhaustive manual labeling process. Expert annotators meticulously evaluated each candidate to assess its capacity to accurately and comprehensively address the corresponding question. Candidates that met stringent criteria for precision, relevance, and completeness were designated as positive samples. These positive samples are intended to exemplify the optimal answers, providing a clear and reliable reference for the model during training. Conversely, any candidate that failed to deliver an effective or accurate answer was classified as a negative sample. This deliberate bifurcation into positive and negative samples ensures that the model is exposed to a distinct contrast between high-quality and substandard responses, thereby enhancing its ability to discriminate between relevant and irrelevant content.

To ensure fairness, consistency, and reproducibility throughout the sample selection process, each question was processed using an identical evaluation protocol. Strict adherence to uniform criteria and standardized methods during the manual labeling phase served to minimize potential biases that might otherwise arise from automated selection procedures. This rigorous approach not only guarantees the high quality of both positive and negative samples but also underpins the overall robustness of the training data. Ultimately, this meticulous curation process is instrumental in bolstering the model’s performance, particularly in its capacity to accurately discern and rank relevant content in response to diverse queries.

\subsection{Training Parameters}

During the fine-tuning of the reranker model, Low-Rank Adaptation (LoRA) was used to reduce the number of parameters that need to be updated during training, thus enhancing training efficiency. Specifically, LoRA was enabled, with the rank set to 32 and the scaling factor set to 64. The learning rate was set to 1e-4 to ensure stable training and prevent gradient explosion. To save memory and improve stability, the batch size per device was set to 1. Additionally, gradient accumulation was enabled, with gradients updated after processing two batches, allowing for more efficient training while maintaining a small batch size. The model was trained using 10 epochs, and mixed precision training was employed with the bf16 setting to further optimize performance and memory usage.

\section{RAG Evaluation}
To evaluate the quality of the Retrieval-Augmented Generation (RAG) model, this study employs two assessment methods: large model evaluation and human evaluation. First, a representative set of questions is collected, covering a range of query scenarios that the model may encounter. For each question, the entire RAG process is executed, including retrieval, generation, and ranking steps, ultimately producing the model’s answers. To facilitate subsequent analysis and evaluation, these generated results are stored in a standardized JSON file format, ensuring traceability and structured data. Through this approach, we can comprehensively assess the performance of the RAG model, evaluating its performance under automated evaluation metrics while also incorporating human review to further analyze the accuracy and relevance of its generated content.

\begin{figure*}
    \centering
    \includegraphics[width=0.8\linewidth]{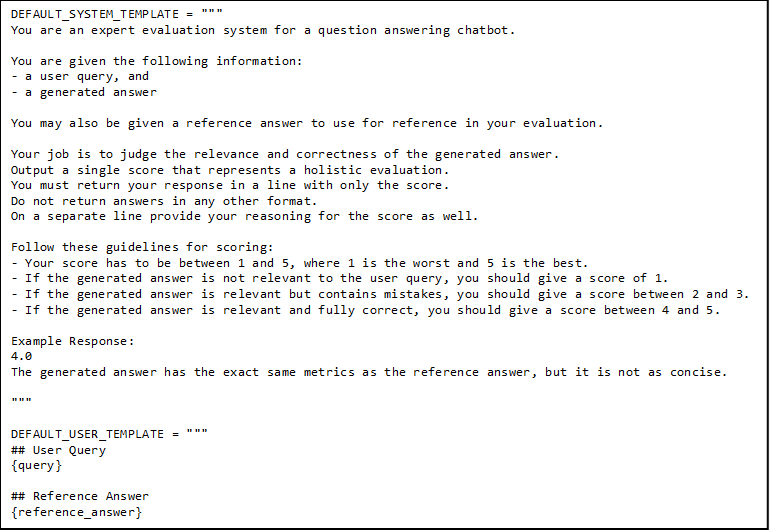}
    \caption{Predefined evaluation prompt template used in LlamaIndex’s \texttt{CorrectnessEvaluator}.}
    \label{fig:evaluation_template}
\end{figure*}

\subsection{Environment Setting}
Our experiments were conducted on a NVIDIA A100-SXM4-80GB server, where we deploy the Qwen2.5-72B-Instruct-AWQ 4bit model. The model was configured using the following command:

\begin{table}[h!]
    \centering
    \caption{\small Model deployment parameters using VLLM.}
    \label{tab:vllm_setting}
    \resizebox{0.4\textwidth}{!}{ % Ensures table fits within half the page width
    \begin{tabular}{p{5cm} c} % Adjusted column widths
        \toprule
        \textbf{Parameter} & \textbf{value} \\
        \midrule
        max-model-len & 6144 \\
        gpu-memoty-utilization & 0.7 \\
        engorce-eager & True \\
        swap-space & 36 \\
        \bottomrule
    \end{tabular}
    }
\end{table}

% \begin{figure}[H]
%     \centering
%     \includegraphics[width=0.9\linewidth]{appendix/vllm.png}
%     \caption{ \texttt{}.}
%     \label{fig:vllm_setip}
% \end{figure}

\subsection{LLM Evaluation}

In this approach, we utilize the open-source tool \textit{llamaindex} ~\citep{Liu_LlamaIndex_2022} to systematically evaluate RAG model outputs by loading two JSON files: one containing generated QA pairs from the RAG model, and the other providing reference answers with contextual data in \texttt{content\_list}. Each question is mapped to its reference information, and a \texttt{Response} object is constructed by combining the RAG-generated answer with its concatenated context. We then employ the \texttt{CorrectnessEvaluator} module (backed by a GPT-4o api) to assess each \texttt{Response}, returning a binary \texttt{passing} flag along with detailed \texttt{feedback}.

To facilitate the evaluation, we integrate the correctness assessment module from LlamaIndex’s \texttt{CorrectnessEvaluator}. This module assigns a numerical score (ranging from 1 to 5) to each response based on alignment with reference answers, using a predefined evaluation template. A threshold-based scoring mechanism determines whether a response meets the required correctness level. The evaluator asynchronously processes input queries, system responses, and reference answers using a structured prompt, ensuring consistency and scalability in judgment.
The predefined evaluation prompt template encapsulated in the module is as follows:

% \begin{figure*}
%     \centering
%     \includegraphics[width=0.8\linewidth]{appendix/prompt_llamaindex.png}
%     \caption{Predefined evaluation prompt template used in LlamaIndex’s \texttt{CorrectnessEvaluator}.}
%     \label{fig:evaluation_template}
% \end{figure*}

During the evaluation, logs and statistics are recorded, including total queries processed, successful evaluations, and skipped items, while exceptions are noted and the evaluation continues for remaining data. Finally, the results—comprising the question, generated answer, reference answer, evaluation status, and feedback—are stored in a structured JSON file for further analysis, providing a robust, LLM-driven metric for measuring and tracing the fidelity of RAG-generated responses.

An example of the result is as below:

\begin{figure*}
    \vspace{0.5cm}
    \centering
    
    \includegraphics[width=0.8\linewidth]{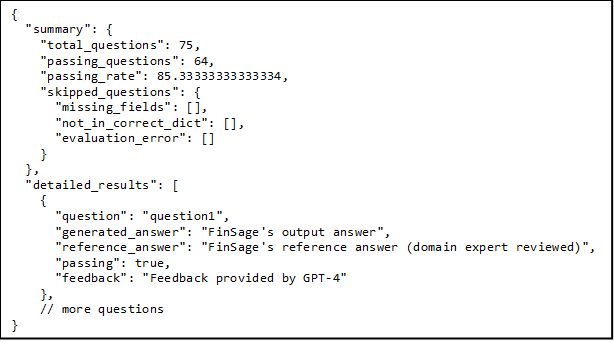}
    \caption{Example question answering results}
    \label{fig:llm_result}
\end{figure*}

\subsection{Re-ranker Evaluation}
\textbf{Binary Normalized Discounted Cumulative Gain (BnDCG)} measures ranking quality by applying position-aware discounting with logarithmic scaling. Since our manually annotated dataset lacks explicit ranking scores for ground truth chunks, we adopt a binary relevance model (1 for relevant, 0 for irrelevant). DCG and its ideal counterpart (IDCG) are computed as:

\begin{equation}
\text{DCG} = \sum_{i=1}^{n} \frac{rel_i}{\log_2(i+1)}
\end{equation}

where \( rel_i \) is the binary relevance score of the retrieved chunk at position \( i \), taking values of either 1 (relevant) or 0 (irrelevant).

The Ideal DCG (IDCG) is computed by sorting all relevant chunks to the top positions:

\begin{equation}
\text{IDCG} = \sum_{i=1}^{n^*} \frac{1}{\log_2(i+1)}
\end{equation}

% where \( n^* \) is the number of relevant chunks in the dataset, ensuring an ideal ranking where all relevant chunks appear first.
where \( n^* \) is the total number of relevant chunks in the dataset, and \( rel_i = 1 \) for all positions \( i \) in the ideal ranking, ensuring that all relevant chunks appear first.
Finally, BnDCG is given by:

\begin{equation}
\text{BnDCG} = \frac{\text{DCG}}{\text{IDCG}}
\end{equation}

ensuring a normalized ranking score between 0 and 1, where 1 represents a perfect ranking.

\begin{table*}
    \centering
    \caption{\small Performance comparison of different re-rankers under different retrieval settings.}
    \label{tab:ranking_comparison_additional}
    \resizebox{0.65\textwidth}{!}{ % Adjust width as needed
    \begin{tabular}{l cccc}
        \toprule
        \textbf{Method} & \textbf{Precision} & \textbf{Normalized Recall} & \textbf{MRR} & \textbf{Binary nDCG} \\
        \midrule
        \multicolumn{5}{l}{\textbf{5 Candidate Results per Retriever (R=5):}} \\
        Top-5 BGE & \Large{0.5913} & \Large{0.6043} & \Large{0.6570} & \Large{0.7540} \\
        Top-5 trained Reranker & \textbf{\Large{0.7324}} & \Large{0.7456} & \textbf{\Large{0.7078}} & \textbf{\Large{0.8124}} \\
        Top-10 BGE & \Large{0.3771} & \Large{0.6881} & \Large{0.3854} & \Large{0.6011} \\
        Top-10 trained Reranker & \Large{0.4424} & \textbf{\Large{0.7783}} & \Large{0.3338} & \Large{0.6097} \\
        \midrule
        \multicolumn{5}{l}{\textbf{10 Candidate Results per Retriever (R=10):}} \\
        Top-5 BGE & \Large{0.6028} & \Large{0.6130} & \Large{0.6540} & \Large{0.7638} \\
        Top-5 trained Reranker & \textbf{\Large{0.7878}} & \Large{0.7910} & \textbf{\Large{0.7545}} & \textbf{\Large{0.8533}} \\
        Top-10 BGE & \Large{0.4133} & \Large{0.5985} & \Large{0.4533} & \Large{0.6285} \\
        Top-10 trained Reranker & \Large{0.5657} & \textbf{\Large{0.8196}} & \Large{0.5155} & \Large{0.6958} \\
        Top-15 BGE & \Large{0.3318} & \Large{0.6889} & \Large{0.3974} & \Large{0.5680} \\
        Top-15 trained Reranker & \Large{0.3883} & \Large{0.8064} & \Large{0.2071} & \Large{0.5354} \\
        \bottomrule
    \end{tabular}
    }
\end{table*}

% \begin{table*}
%     \centering
%     \caption{\small Performance of End-To-End Answer (ROUGE scores included)}
%     \label{tab:resposne_comparison_additional}
%     \resizebox{0.65\textwidth}{!}{ % Ensures table fits within half the page width
%     \begin{tabular}{p{4cm} ccccc} % Adjusted column widths
%         \toprule
%         \textbf{Dataset/Method} & \textbf{LLM} & \textbf{Mannual} & \textbf{ROUGE1} & \textbf{ROUGE2} & \textbf{ROUGEL} \\
%         \midrule
%         FinanceBench/FinSage & \Large{0.4966} & \Large{0.5705} & \Large{0.3319} & \Large{0.1542} & \Large{0.3002} \\
%         Company/FinSage & \Large{0.8533} & \Large{0.8800} & \Large{0.7213} & \Large{0.5576} & \Large{0.6657} \\
%         \bottomrule
%     \end{tabular}
%     }
% \end{table*}

\begin{figure*}
    \includegraphics[width=0.97\textwidth]{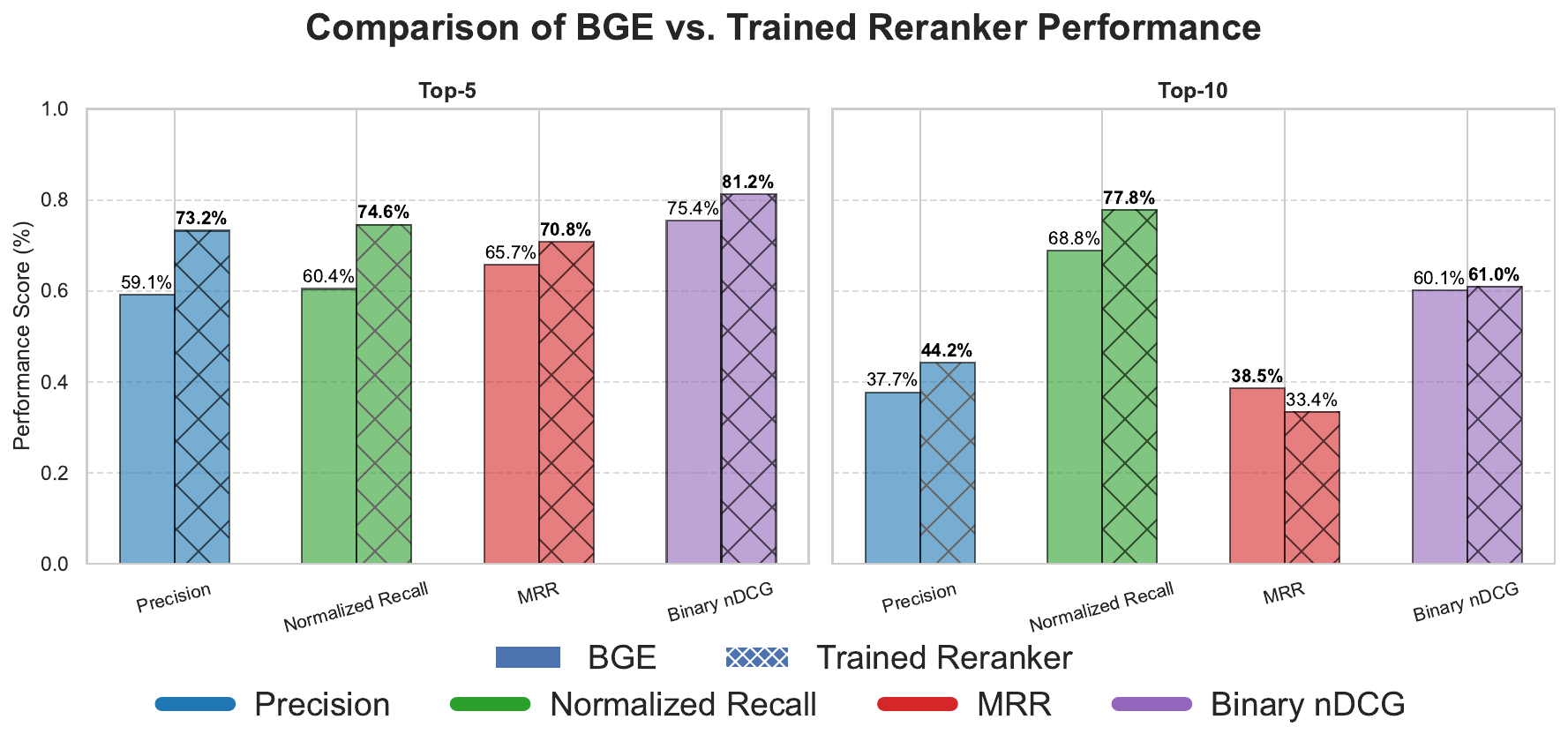} \caption{\small Performance comparison of different re-rankers under different retrieval settings when limiting the retrieval bundle to 5.}
    \label{fig:ranking_comparison_appendix}
\end{figure*}

\section{Efficiency and cost analysis}
% System Efficiency and Deployment Cost Table
\begin{table*}
    \centering
    \caption{System Efficiency and Deployment Cost}
    \label{tab:system_efficiency}
    \resizebox{0.9\textwidth}{!}{
    \begin{tabular}{clcccc}
        \toprule
        \textbf{Step} & \textbf{Process} & \textbf{Time (s)} & \textbf{Model Used} & \textbf{Token Usage} & \textbf{Cost Estimation} \\
        \midrule
        1 & Query rewriting & $\sim$2.5 & GPT-4o & $\sim$1200 & $\sim$\$0.005 \\
        2 & HyDE per sub-query (async) & $\sim$4.2 & GPT-4o & $\sim$500 & $\sim$\$0.002 $\times$ $n$ \\
        3 & Retrieval \& reranking & $\sim$4.7 & Local & N/A & 24GB GPU \\
        4 & Sub-query answering (async) & $\sim$4.7 & GPT-4o & $\sim$2500 & $\sim$\$0.012 $\times$ $n$ \\
        5 & Final answer merging & $\sim$1.7 & GPT-4o & $\sim$200 + 200 $\times$ $n$ & $\sim$\$0.002 + \$0.002 $\times$ $n$ \\
        \midrule
        \textbf{Total} & --- & \textbf{$\sim$13--17} & --- & $\sim$3.7k + 3k $\times$ $n$ & \textbf{$\sim$\$0.017 + \$0.016 $\times$ $n$} \\
        \bottomrule
    \end{tabular}
    }
    \caption*{$n$ denotes the number of sub-queries.}
\end{table*}

% Retrieval Latency Table
\begin{table}[h!]
    \centering
    \caption{Retrieval Latency.}
    \label{tab:retrieval_latency}
    \resizebox{0.3\textwidth}{!}{
    \begin{tabular}{p{2cm} c}
        \toprule
        \textbf{Method} & \textbf{Avg Time (s)} \\
        \midrule
        FAISS & 0.057 \\
        Metadata(FAISS) & 0.050 \\
        BM25 & 0.014 \\
        \midrule
        \textbf{Total} & \textbf{0.121} \\
        \bottomrule
    \end{tabular}
    }
\end{table}

To evaluate computational efficiency, we conducted detailed benchmarks using an NVIDIA H20 GPU. This hardware surpasses the minimum system requirement of a GPU with 24GB of memory, making it available to increase the batch size for the re-ranker model, resulting in higher inference throughput. Table \ref{tab:system_efficiency} summarizes key performance metrics, specifically the latency and estimated cost per user request. Latency is defined as the time elapsed from receiving a user request to generating the first token of the final response. Our analysis indicates that the primary computational costs originate from two main sources: the LLM inference steps, where latency scales with the length of the generated output, and the re-ranker model inference. And the estimated average cost to process a request involving a single sub-query is approximately 13 seconds with \$0.03. Furthermore, Table \ref{tab:retrieval_latency} details the retrieval latency for the In-depth company dataset, demonstrating the efficiency of our multi-path retriever implementation.

\section{End-user feedback analysis}
Our system is actively used by a major industrial client, CG Matrix Technology Limited, where FinSage has processed 2,702 user queries with a strong average satisfaction score of 4.19/5.0. The satisfaction score is the rating users give to each response generated by \ourmodel{}, with scores ranging from 1-5 (including half-point ratings like 4.5). Overall score distribution in table \ref{tab:score_distribution} reveals that 81.87\% of all queries received a score of between 4 and 5, whcih indicates high end-user satisfaction with \ourmodel{}'s responses. 

As shown in table \ref{tab:question_types}, we also categorized the questions into RAG questions (91.9\%) and Non-RAG questions (8.1\%). The distribution indicates that the predominant use of the system focuses on retrieving domain-specific financial information, which aligns with our intended design goals. The difference of average satisfaction score between RAG (4.13/5.0) and Non-RAG questions (4.80/5.0) derives from the inherent complexity of financial queries requiring precise retrieved context versus more general knowledge questions that LLM that already acquired.

For RAG questions, we categorize them into 4 categories, each addressing distinct informational needs in the financial domain as shown in table \ref{tab:rag_categories}:
\begin{itemize}
\item \textbf{Business, Products, Competition (35.2\%)}: This category encompasses queries about product lines, pricing strategies, sales channels, market positioning, and competitive advantages. Users frequently inquire about specific product offerings, international market presence, competitive landscape analysis, research and development investments, and strategic business advantages. These questions require the system to synthesize information across product specifications, market analysis documents, and competitive intelligence reports.
\item \textbf{Basic Information, Equity Structure (28.3\%)}: Questions in this category focus on fundamental company information, stock data, shareholder composition, management profiles, and corporate governance details. Users seek information about headquarters location, stock exchange listings, major shareholders, board composition, and corporate relationships. Answering these queries requires precise extraction of factual data points from corporate filings, annual reports, and governance documents.

\item \textbf{Financial Status, Performance (25.8\%)}: This category contains queries about financial metrics, operational performance, profitability measures, and forward-looking projections. Users typically request specific financial indicators such as gross margins, operational profits/losses, delivery volumes, sales forecasts, and financial ratios. These questions often involve temporal reasoning and numerical analysis across multiple financial statements and reports.

\item \textbf{Regulatory Policy (2.5\%)}: The smallest but highly specialized category addresses regulatory frameworks, tariff structures, geopolitical influences, and policy impacts. Users inquire about tariff applications, subsidy programs, tax incentives, and regulatory compliance requirements. These complex queries require the system to interpret regulatory documents and assess their implications on business operations.
\end{itemize}

While examining specific RAG question categories (table \ref{tab:rag_categories}), we notice \textbf{Business, Products, Competition} queries were most frequent (35.2\%) with medium average satisfication score, while for \textbf{Regulatory, Policy} questions, even it represents only 2.5\% of total queries, it receives relatively high satisfaction scores (4.40/5.0), which demonstrates \ourmodel{}'s effectiveness in handling complex compliance-related inquiries. However, \textbf{Financial status, Performance} queries receive the lowest average satisfaction score (3.88/5.0) among RAG categories, this suggests the area for potential improvement, which requires more precise numerical data retrieval and temporal reasoning across multiple chunks that \ourmodel{} might have to incoporate.

% Table: Detailed Score distribution
\begin{table}[h!]
    %\centering
    \caption{\small Detailed Score Distribution}
    \label{tab:score_distribution}
    \resizebox{0.48\textwidth}{!}{
    \begin{tabular}{p{3cm} cc}
        \toprule
        \textbf{\small Score} & \textbf{\small Count} & \textbf{\small Percentage} \\
        \midrule
        \small 1.0 & \small 100 & \small 3.70\% \\
        \small 2.0 & \small 76 & \small 2.81\% \\
        \small 2.5 & \small 33 & \small 1.22\% \\
        \small 3.0 & \small 206 & \small 7.62\% \\
        \small 3.5 & \small 75 & \small 2.78\% \\
        \small 4.0 & \small 227 & \small 8.40\% \\
        \small 4.5 & \small 1466 & \small 54.26\% \\
        \small 5.0 & \small 519 & \small 19.21\% \\
        \bottomrule
    \end{tabular}
    }
\end{table}

% Table： Question Types
\begin{table}[h!]
    \centering
    \caption{\small Distribution and Performance of RAG vs Non-RAG Questions}
    \label{tab:question_types}
    \resizebox{0.48\textwidth}{!}{
    \begin{tabular}{p{3.5cm} ccc}
        \toprule
        \textbf{\small Question Type} & \textbf{\small Count} & \textbf{\small Percentage} & \textbf{\small Avg. Score} \\
        \midrule
        \small RAG questions & \small 2,482 & \small 91.9\% & \small 4.13/5.0 \\
        \small Non-RAG questions & \small 220 & \small 8.1\% & \small 4.80/5.0 \\
        \bottomrule
    \end{tabular}
    }
\end{table}

% Table: RAG Question Categories
\begin{table*}
    \centering
    \caption{\small Distribution and Performance of user feedback by RAG Question Category. Percentages shown are relative to all RAG questions, which constitute a subset of total questions in the user queries.}
    \label{tab:rag_categories}
    \resizebox{0.65\textwidth}{!}{
    \begin{tabular}{p{4cm} rrr}
        \toprule
        \textbf{RAG Question Category} & \textbf{Count} & \textbf{Percentage} & \textbf{Avg. Score} \\
        \midrule
        business\_products\_competition & 952 & 38.4\% & 4.23/5.0 \\
        basic\_info\_equity\_structure & 765 & 30.8\% & 4.22/5.0 \\
        financial\_status\_Performance & 697 & 28.1\% & 3.88/5.0 \\
        regulatory\_policy & 68 & 2.7\% & 4.40/5.0 \\
        \bottomrule
    \end{tabular}
    }
\end{table*}

% \begin{table}
%     \centering
%     \caption{Response Time Comparison (in seconds)}
%     \label{tab:graph_light_rag}
%     \resizebox{0.4\textwidth}{!}{
%     \begin{tabular}{p{3cm} cccc}
%         \toprule
%         \textbf{Method} & \textbf{Mean} & \textbf{Median} & \textbf{Min} & \textbf{Max} \\
%         \midrule
%         GraphRAG & 16.90 & 15.27 & 9.86 & 40.67 \\
%         LightRAG & 12.16 & 8.84 & 3.41 & 859.51 \\
%         FinSage & 19.34 & 18.57 & 8.57 & 40.02 \\
%         \bottomrule
%     \end{tabular}
%     }
% \end{table}

% \begin{table}
%     \centering
%     \caption{Faithful Evaluation Score Comparison}
%     \label{tab:graph_light_rag_faithful_score}
%     \resizebox{0.48\textwidth}{!}{
%     \begin{tabular}{p{1.5cm} cccccc}
%         \toprule
%         \textbf{Method} & \textbf{Questions} & \textbf{Mean} & \textbf{Median} & \textbf{Min} & \textbf{Max} & \textbf{Pass \%} \\
%         \midrule
%         GraphRAG & 71(w/ 4 failures) & 3.46 & 3.50 & 2.0 & 5.0 & 42.50\% \\
%         LightRAG & 75 & 2.45 & 2.00 & 1.0 & 5.0 & 13.67\% \\
%         FinSage & 75 & 4.31 & 5.00 & 1.00 & 5.00 & 82.67\% \\
%         \bottomrule
%     \end{tabular}
%     }
% \end{table}

% \section{Config of GraphRAG and LightRAG}
\section{Alternative Methodology comparison}~\label{graphrag}
We also explore the usage of graph-based representations, we conducte additional experiments with \textbf{GraphRAG} and \textbf{LightRAG} due to their mainstream usage for graph-based retrieval, showing unsatisfactory performance from both graph-based retrieval models (Table \ref{tab:graph_light_rag} and \ref{tab:graph_light_rag_faithful_score}).
% despite training and testing under the similar setup 

\subsection{GraphRAG Configuration}
GraphRAG is a graph-based retrieval system that uses knowledge graph representations to enhance contextual understanding. The configuration is detailed in Table \ref{tab:graphrag_config}. 

\begin{table}[h!]
\caption{GraphRAG Configuration Parameters and Methods}
\label{tab:graphrag_config}

\begin{tabular}{|p{3.8cm}|p{4cm}|}
\hline
\textbf{Parameter / Method} & \textbf{Value / Description} \\ 
\hline
% --- Model Configuration ---
\multicolumn{2}{|c|}{\textit{Chat Model}} \\
\hline
Model & gpt-4o \\
Concurrent Requests & 25 \\
Async Mode & threaded \\
\hline
\multicolumn{2}{|c|}{\textit{Embedding Model}} \\
\hline
Model & text-embedding-3-small \\
Concurrent Requests & 25 \\
\hline
% --- Data Processing ---
\multicolumn{2}{|c|}{\textit{Data Processing}} \\
\hline
Chunk Size & 1200 \\
Chunk Overlap & 100 \\
Group By & id \\
Vector Store & lancedb \\
\hline
% --- Graph Processing ---
\multicolumn{2}{|c|}{\textit{Graph Processing}} \\
\hline
Entity Types & organization, person, geo, event \\
Extract Claims & disabled (optional) \\
Max Cluster Size & 10 \\
Summary Max Length & 500 \\
\hline
% --- Query Methods ---
\multicolumn{2}{|c|}{\textit{Query Methods}} \\
\hline
local\_search & Context-dependent information \\
global\_search & Map-reduce approach \\
drift\_search & Handles concept drift \\
basic\_search & Simple retrieval mechanism \\
\hline
\end{tabular}
\end{table}

\subsection{LightRAG Configuration}
LightRAG represents a more streamlined approach which prioritizes efficiency while maintaining retrieval quality. The detailed configuration is shown in Table \ref{tab:lightrag_config}.

\begin{table*}[htbp!] % Use placement specifiers like htbp (Here, Top, Bottom, Page)
\caption{LightRAG Configuration Parameters}
\label{tab:lightrag_config}

\begin{tabular}{|p{3.8cm}|p{4cm}|} % Adjust column widths as needed
\hline
\textbf{Parameter} & \textbf{Value / Description} \\
\hline
% --- Language Model ---
\multicolumn{2}{|c|}{\textit{Language Model}} \\
\hline
Model & GPT-4o \\
Configuration & custom\_llm\_model\_func \\
\hline
% --- Embedding Model ---
\multicolumn{2}{|c|}{\textit{Embedding Model}} \\
\hline
Model & text-embedding-3-small \\
Dimension & 1536 (auto-detected) \\
Max Token Size & 8192 \\
\hline
% --- Parallel Processing ---
\multicolumn{2}{|c|}{\textit{Parallel Processing}} \\
\hline
embedding\_batch\_num & 128 \\
embedding\_func\_max\_async & 64 \\
max\_parallel\_insert & 8 \\
llm\_model\_max\_async & 16 \\
insert\_batch\_size & 32 \\
\hline
% --- Query Parameters ---
\multicolumn{2}{|c|}{\textit{Query Parameters - Retrieval Modes}} \\
\hline
mode & mix (default), local, global, hybrid, naive \\
top\_k & 60 (default, configurable) \\
\hline
\multicolumn{2}{|c|}{\textit{Query Parameters - Token Limits}} \\
\hline
max\_token\_for\_text\_unit & 4000 \\
max\_token\_for\_global\_context & 4000 \\
max\_token\_for\_local\_context & 4000 \\
\hline
% --- Response Controls ---
\multicolumn{2}{|c|}{\textit{Response Controls}} \\
\hline
response\_type & "Multiple Paragraphs" \\
stream & false \\
only\_need\_context & false \\
only\_need\_prompt & false \\
\hline
% --- Keyword Parameters ---
\multicolumn{2}{|c|}{\textit{Keyword Parameters}} \\
\hline
hl\_keywords & High level keywords: retrieval prioritization \\
ll\_keywords & Low level keywords: refinement \\
\hline
\end{tabular}
\end{table*}

\subsection{Comparative Analysis}
Our experiment results presented in Tables \ref{tab:graph_light_rag} and \ref{tab:graph_light_rag_faithful_score} reveal the important insights regarding the performance trade-offs between GraphRAG and LightRAG implementations, compared to \ourmodel{}. 

While both retrieval models do achieve enhanced contextual understanding through structural relationships, we notice their unsatisfactory performance across several key metrics. \textbf{GraphRAG} achieved moderate faithful evaluation scores (mean: 3.46) with a 42.50\% pass rate but at the cost of longer response times (mean: 16.09s). \textbf{LightRAG} optimized for speed (mean: 12.16s) but delivered the poorest faithful evaluation performance with only a 13.67\% pass rate, which is not acceptable in the financial question answering domain. In general, \ourmodel{} demonstrates superior faithful evaluation scores (mean: 4.31) with an impressive 82.67\% pass rate, although with slightly longer response times (mean: 19.34s). This suggests that the additional complexity introduced by graph-based representation does not translate to improved response quality in our experimental context. These findings emphasize the importance of balancing response speed with response quality when selecting RAG implementations for specific application domains.

% \begin{table}[h!]
%     \centering
%     \caption{\small Response Time Comparison (in seconds)}
%     \label{tab:graph_light_rag}
%     \resizebox{0.48\textwidth}{!}{
%     \begin{tabular}{p{3cm} cccc}
%         \toprule
%         \textbf{\small Method} & \textbf{\small Mean} & \textbf{\small Median} & \textbf{\small Min} & \textbf{\small Max} \\
%         \midrule
%         \small GraphRAG & \small 16.90 & \small 15.27 & \small 9.86 & \small 40.67 \\
%         \small LightRAG & \small 12.16 & \small 8.84 & \small 3.41 & \small 859.51 \\
%         \small FinSage & \small 19.34 & \small 18.57 & \small 8.57 & \small 40.02 \\
%         \bottomrule
%     \end{tabular}
%     }
% \end{table}

% 5 Candidate Results per Retriever

\section{Additional Results}
\subsection{Re-ranker Results}
We also assess the impact of Document Re-ranker on retrieval effectiveness using different $R$, as detailed in Table \ref{tab:ranking_comparison_additional}. $R$ denotes the number of candidate bundles for each retriever. The retrieval system under this assessment integrates four retrievers—FAISS (baseline), BM25, Metadata, and HyDE—retrieving at least $4\times R$ chunks. While increasing $R$ enhances the chances of retrieving relevant chunks, it also intensifies the re-ranker’s challenge of filtering out irrelevant ones.

% \subsection{Final Response}
% As part of our evaluation, we also compute standard metrics as shown in Table \ref{tab:resposne_comparison_additional}, including ROUGE1, ROUGE2 and ROUGE-L scores, to assess the alignment between system-generated responses and reference answers. However, these scores often fall short in capturing the semantic equivalence of different responses, which leads to potentially misleading evaluations. Given such limitation, we rely primarily on LLM-based evaluation using GPT-4o and human labeling,

\end{document}